\documentclass[9pt,twoside]{extarticle}

\usepackage[margin=0.75in, top=0.75in, bottom=0.75in,
            columnsep=0.25in]{geometry}
\usepackage{multicol}

\usepackage[T1]{fontenc}
\usepackage{mathptmx}
\usepackage[scaled=0.86]{helvet}
\usepackage{microtype}

\usepackage{amsmath,amssymb,bm}
\usepackage{mathtools}

\usepackage{float}

\usepackage{graphicx}
\usepackage{booktabs}
\usepackage{dcolumn}
\usepackage[rightcaption]{sidecap}
\usepackage[labelfont=bf, 
            font=small,
            labelsep=period]{caption}
\usepackage{enumitem}

\usepackage{xcolor}
\usepackage{url}
\newcommand{\doi}[1]{\href{https://doi.org/#1}{\detokenize{#1}}}
\usepackage{hyperref}
\usepackage{hycolor}
\hypersetup{colorlinks=true,citecolor=[rgb]{0,0.18,0.68}, 
urlcolor=[rgb]{0,0.18,0.68}, linkcolor=[rgb]{0,0.18,0.68}, final = true}
\renewcommand{\url}[1]{\href{#1}{\detokenize{#1}}}
\usepackage{relsize}

\usepackage[super,compress,sort]{natbib}
\setcitestyle{numbers,super,open={},close={}}

\usepackage{titlesec}
\titleformat{\section}{\small\bfseries\sffamily\uppercase}{\thesection.}{0.5em}{}
\titleformat{\subsection}{\small\bfseries\sffamily}{\thesubsection.}{0.5em}{}
\titleformat{\subsubsection}{\small\itshape}{\thesubsubsection.}{0.5em}{}

\usepackage{fancyhdr}
\usepackage{abstract}

\usepackage{kotex}

\usepackage{siunitx}
\titleformat{\section}
    {\fontsize{12}{10}\bfseries\sffamily\uppercase}
    {\thesection.}{0.5em}{}

  \def\be{\begin{equation*}}
  \def\ee{\end{equation*}}
  \def\ba{\begin{eqnarray}}
  \def\ea{\end{eqnarray}}
  
  \def\fref#1{Fig.~\ref{#1}}

  \def\bt{\textrm} %\textsf
  
  \def\nsb#1{\noindent\textbf{\bt{#1~}}}
  \def\nsi#1{\noindent\textit{\bt{#1~--}}}
  \definecolor{or}{RGB}{234,142,53}
  \definecolor{gr}{RGB}{150,150,150}
  \definecolor{bl}{RGB}{54,152,187}

  \newcommand{\ie}{\textit{i.e.}}
  \newcommand{\eg}{\textit{e.g.}}

  \definecolor{YKB}{rgb}{0.00,0.18,0.65}

  \def\Nnotes{N}
  \def\seqMidi{\mathbf{p}}
  \def\seqChroma{\mathbf{c}}
  \def\seqTmidi{\seqMidi^\tau}
  \def\seqTchroma{\seqChroma^\tau}
  \def\seqDur{\mathbf{d}}

  \def\vecMidi{\mathbf{x}}
  
  \def\vecTmidi{\mathbf{x}_\tau}
  \def\vecTchroma{\mathbf{x}_{c,\tau}}

  \def\nmeas{N_{\textrm{measure}}}
  \def\pid{\textrm{PID}}

\begin{document}

\title{\LARGE\bfseries\sffamily Contrasting statistical patterns in melodic and
 molecular evolution reveal distinctive constraints in a culturally evolving system.}

\author{
    John M. McBride\textsuperscript{1,*} 
    \and 
    W. Tecumseh Fitch\textsuperscript{1,†}
}

\date{
    \small
    \textsuperscript{1}Department of Behavioral and Cognitive Biology, 
    University of Vienna, Vienna, Austria\\[0.5em]
    \textsuperscript{*}Correspondence: 
    \href{mailto:jmmcbride@protonmail.com}{jmmcbride@protonmail.com}\\
    \textsuperscript{†}Correspondence: 
    \href{mailto:tecumseh.fitch@univie.ac.at}{tecumseh.fitch@univie.ac.at}
}

\twocolumn[
  \begin{@twocolumnfalse}
    \maketitle
  \end{@twocolumnfalse}
]

\begin{abstract}
  Evolved sequences can be used to infer the rules of evolution.
  Orally transmitted folk melodies are evolved sequences whose similarity to
  protein sequences (one-dimensional, drawn from a limited alphabet)
  invites application of bioinformatics methods to study cultural evolution.
  A major obstacle is that melodies encode rhythm, which breaks some assumptions
  of standard sequence-alignment algorithms. We develop a rhythm-aware alignment
  method and apply it to \num{40000} Irish dance tune variants, enabling the first
  large-scale automated melodic alignment.
  Four canonical bioinformatics analyses -- mutability, substitution matrices,
  positional conservation, and covariance -- reveal patterns distinct from those
  of molecular evolution, revealing the forces that shape each domain:
  biochemical and biophysical constraints for proteins;
  memory, motor, and social biases for melodies.
  Together the results show that bioinformatics provides a powerful framework
  -- conceptual as much as algorithmic -- for studying cultural evolution.
  Although the cultural transmission of music has been discussed for centuries,
  here we show how to analyze it at large scale.
\end{abstract}
\vspace{0.5cm}
\vspace{0.5cm}

\section*{Introduction} 

  Sequences generated by evolutionary processes are a wellspring
  of information that can be used to infer the rules by which they
  evolved. Examples from language span the decoding of ancient languages,\cite{braovicSystematic2024}
  to the development of modern language models.\cite{wangHistory2025}
  Analysis of biological sequences allows us to infer evolutionary histories,\cite{felsensteinInferring2004,mifsudRecent2025}
  the rules of protein biophysics and evolution,\cite{echaveCauses2016,orenbuchProteomewide2025}
  and to predict future evolutionary trends.\cite{huotPredicting2025,hamelinPredicting2025}
  Melodies are also sequences that have been shaped and selected through cultural evolution,
  \cite{sharpEnglish1907,bayardProlegomena1950,janMemetics2007,savageCultural2019a,youngbloodCultural2023a}
  and hold untapped potential for understanding the rules and history of musical evolution,
  and thus perhaps of cultural evolution more generally.

  Algorithms offer a clear route to knowledge transfer from bioinformatics. Melodies
  can be represented as 1-dimensional pitch sequences, and sequence alignment tools
  can in principle be used to compare melodies directly.\cite{savageSequence2022a,hajicjrBuilding2023}
  However, there are noteworthy distinctions both in the form and the evolution
  of melodic and biological sequences.
  Letters in biological sequences always represent the same chemical unit, while
  for melodies, notes have two effective dimensions -- pitch and rhythm.
  When notes are subdivided into multiple shorter notes, or fused into longer notes,
  alignment algorithms treat this as a `gap', or an insertion/deletion
  (indel),\cite{needlemanGeneral1970} but this is qualitatively different from
  the insertions and deletions that arise in biological sequences. 
  These distinctions can at best lead to suboptimal performance
  when using bioinformatics algorithms out-of-the-box, but could
  easily lead to major systematic errors.

  Knowledge transfer from bioinformatics is not limited to algorithms,
  but also broader methodological approaches to discovering the statistical
  regularities in sets of sequences. Key analyses such as amino acid mutability, substitution matrices,
  sequence conservation and covariance, all led to important discoveries
  and are foundational to modern bioinformatic analysis.
  \cite{dayhoff221978,henikoffAmino1992,trivediSubstitution2020,huangNo2006,leePredicting2007}
  For example, sequence covariance -- correlations between substitutions at different positions --
  can inform us about how a protein folds,\cite{hopfSequence2014} such that we
  can now predict structure given sufficient co-evolutionary information.\cite{jumna21,baesc21}
  Here, we propose that bioinformatics analyses can be tailored to
  melodies, as detailed below, to uncover general rules underlying melodic evolution.

  As a case study in how to adapt bioinformatics methods for melodies we focus
  on Irish dance tunes. We chose this tradition for three reasons.
  Folk tunes in general have the right evolutionary properties,
  dance tunes in particular are methodologically tractable,
  and Irish music is uniquely well-documented.
  We now elaborate on each of these reasons.

  First, folk tunes persist in recognizable forms over long periods, change to produce variants,
  and are subject to cultural selection.\cite{karpelesDefinition1955,cowderyMelodic1990}
  Change comes from transmission by ear, which introduces copying errors through limits
  on perception and memory.\cite{breathnachUse1986,janMemetics2007}
  Intentional variation adds further change, as melodic improvisation is a valued skill
  in Irish folk music.\cite{ohallmhurainShort2017}
  These forces -- copying fidelity, conformity bias from ensemble playing, and novelty bias
  from individual expression -- produce an ecosystem of tunes and variants, analogous
  to genes and their homologues, which can be grouped into tune families based
  on shared melodic history.\cite{creanzaCultural2017,bayardProlegomena1950}

  Second, dance tunes are especially tractable for rhythm-aware sequence analysis.
  As they are designed for dancing, they obey strict metrical and
  structural constraints:\cite{osuilleabhainCreative1990,dohertyMelodic2022} parts have fixed lengths,
  contain no rests, and can therefore be aligned without gaps -- sidestepping
  one of the central difficulties of sequence alignment. Tunes also conform
  to specific tonalities, with pitches drawn from sets of seven or fewer pitch classes.

\begin{figure*}[t]
\centering
\includegraphics[width=0.95\textwidth]{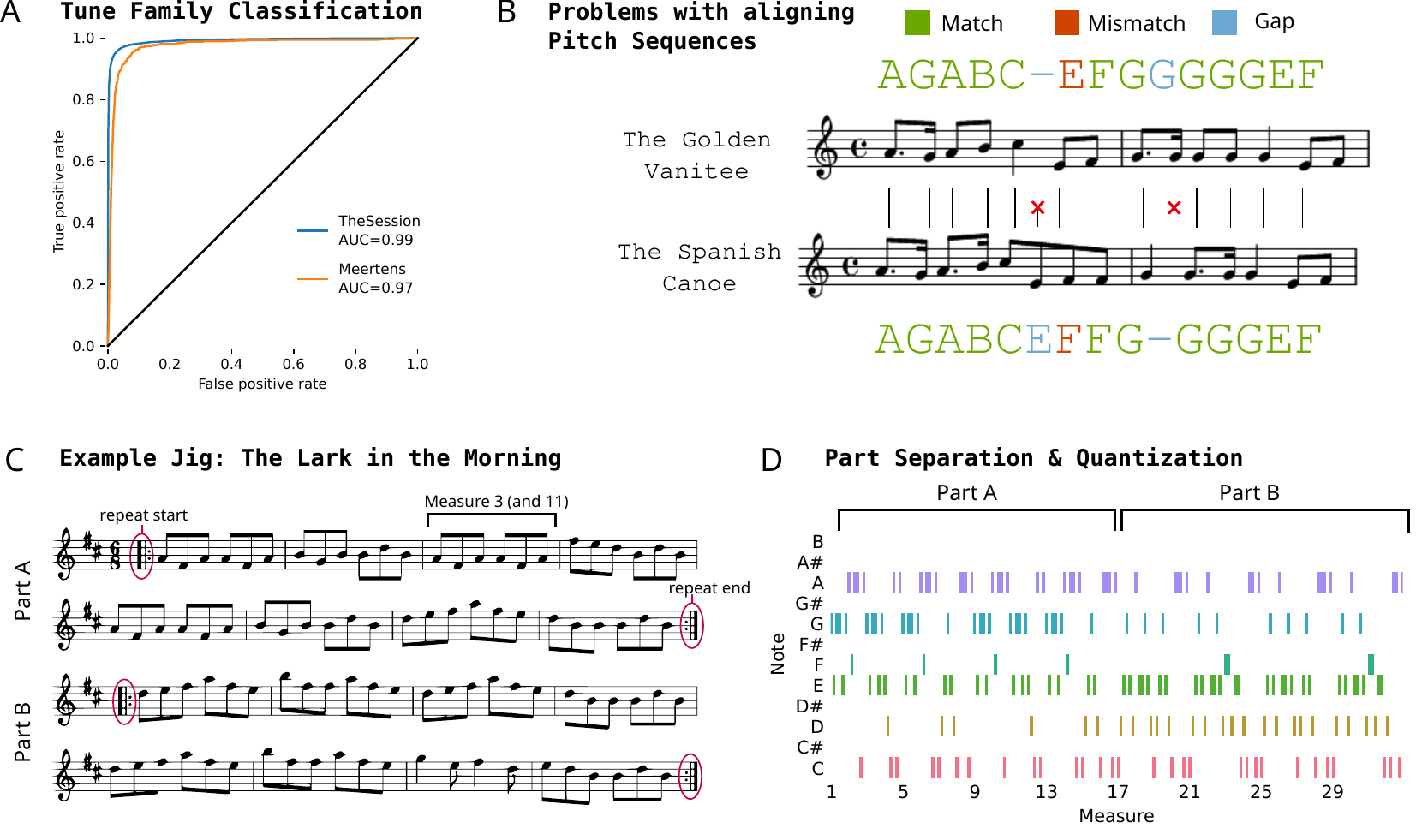}
\caption{\label{fig:fig1}
  \textbf{Pitch-only and rhythm-aware alignments.}
  \textbf{(A)}~Receiver operating curve (ROC) and area-under-the-curve (AUC) values for classifying
  tunes into families using pitch sequences.
  \textbf{(B)}~Illustration of two problems with aligning melodies without rhythm information.
  We show excerpts of two melodies from the Bronson dataset, alongside the
  manually-corrected pitch sequence alignments.\cite{savageSequence2022a}
  Without knowledge of rhythm, the alignment to the 'EFF' part of the lower sequence
  will always be 'E-F' or 'EF-', instead of the correct alignment, '-EF'.
  The region with 'GGGGG' matched to 'G-GGG' is not a unique solution. Other top-scoring
  alignment solutions include '-GGGG', 'GG-GG', 'GGG-G', 'GGGG-'.
  \textbf{(C)}~An example of a jig (6/8 meter). These are the first two parts of the four-part jig,
  ``The Lark in the Morning''. Repeat lines are highlighted with red circles. After expanding
  repeats there are 32 measures.
  \textbf{(D)}~The same jig (C) represented in the format used for alignments. To get this we expand repeats,
  map the notes onto a grid with a standard grid size (in this case, eighth notes), and separate tunes into parts.
}
\end{figure*}

  Finally, Irish dance tunes stand out for the richness and scale of available data.
  Multiple dance forms and tonalities produce structured variation that
  is both diverse and constrained -- an ideal combination for statistical analysis.
  This tradition is also unusually well-documented online, owing in part to the
  ABC notation format developed in 1994,\cite{walshawAnnouncing1994} which made
  sharing tune transcriptions easy and led to repositories including (but not limited to) thesession.org,
  which holds over \num{50000} community-contributed tunes.\cite{keithThesessionorg,kuntzTunearchorg,ngIrishtuneinfo,walshawAbcnotationcom}
  The total number of tunes and variants available in digital form online exceeds \num{100000}, a uniquely large dataset.

  The concept of tune evolution and tune families has been developed over many years,
  \cite{sharpEnglish1907,bayardProlegomena1950,bronsonBallad1969,cowderyFresh1984}
  but empirical research on how melodies evolve remains in its infancy.
  \cite{savageSequence2022a,hajicjrBuilding2023,lebominEvolution2016,streetRole2022,nishikawaCultural2022a,nishikawaExploring2025}
  Previous efforts involving manual comparison of similar tunes,\cite{bayardTwo1954,cowderyMelodic1990,grassoMelodic2011,janssenPredicting2017a}
  are not scalable to large datasets. As for computational approaches,
  most have been dealing with orthogonal topics such as tune family identification,
  \cite{vankranenburgMusical2009,lavinSimilarity2010,savageAutomatic2015,bountouridisMelodic2017,diamondAutomatic2024}
  or phylogeny.\cite{windramPhylogenetic2014,juhaszSimultaneous2019,hajicjrGenome2025}
 In a seminal contribution, \citet{savageSequence2022a} used sequence analysis tools for similar tune identification.
  They did attempt using bioinformatics tools to align sequences, but due to poor quality of automated alignments 
  had to manually align their sequences. Thus they compared only \num{328} pairs of tunes, several orders
  of magnitude lower than the number in principle permitted for a joint collection numbering
  \num{10000} melodies. Furthermore, the melodies were encoded in a reduced format
  that excluded most rhythmic, and some melodic information.
  Thus, compared to what has been discovered in biology or linguistics,
  existing studies on melody evolution have only scratched the surface of what is possible in theory.

  Here, we first show that bioinformatics algorithms are useful for identifying similar tune
  variants, but not for aligning them. We develop a novel rhythm-aware alignment method
  suitable for musical traditions with rigid metrical structure.
  Applying it to \num{40000} Irish dance tune variants, we run four canonical
  bioinformatics analyses (mutability, substitution matrices, 
  positional conservation, and covariance)
  and show that the resulting statistical signatures highlight clear differences between molecular
  and melodic sequence evolution. We then use the results to devise hypotheses and analyses
  designed to uncover the constraints that guide melodic evolutionary paths.
  The analyses reveal statistical signatures that are categorically
  different from those of protein evolution. In particular, we find that pitch mutability tracks
  tonal hierarchy, reflecting memory and perceptual constraints on transmission;
  substitution rates mirror the melodic interval distribution, reflecting
  motor constraints on melodic motion; positional conservation correlates with
  metrical hierarchy, and covariance reveals conserved repetition structure.
  The last two findings are consistent with the long-standing hypothesis that
  a stable ``melodic skeleton'' encodes tune identity\cite{osuilleabhainCreative1990},
  and we outline experimental predictions that would test it.

\section*{Results} 

\nsb{Is pitch sufficient to identify similar melodies?}
  Do pitch sequences, without any rhythm or note durations, contain sufficient information
  for identifying similar melodies?
  If you attend an Irish traditional music session you can see participants,
  if they do not know the tune that is being played, using an app -- TunePal,\cite{dugganTunepal2011a}
  or FolkFriend\cite{wyllieFolkFriend} -- to query the tune.
  These apps work well, but their algorithms use both pitch and rhythm.\cite{dugganSystem2008}
  Whether pitch alone is sufficient has been sporadically investigated with no clear consensus.
  One study found moderate success on \num{6000} Dutch songs,\cite{vankranenburgMusical2009}
  while others used small sample sizes\cite{savageAutomatic2015,bountouridisMelodic2017}
  or non-standard evaluation metrics.\cite{savageSequence2022a,vankranenburgCrossCorpus2023}

  We tested this using the standard algorithm MMseqs2\cite{steineggerMMseqs22017}
  on two large datasets (TheSession, Meertens),
  amounting to billions of tune comparisons. The algorithm first eliminates pairs that are
  unlikely to match, then computes percent identity ($\pid$) scores -- a measure
  of sequence similarity -- for the remainder using local alignment.
  The overall true positive rates -- including pairs eliminated at the pre-filtering stage --
  were 0.71 (TheSession) and 0.42 (Meertens).
  Thus pitch alignment alone, deprived of rhythm information, misses a substantial fraction of related tunes,
  particularly those that have diverged considerably.

  Among the pairs that do pass pre-filtering, however, $\pid$ is a highly accurate classifier.
  ROC curves (\fref{fig:fig1}A) show AUC values of 0.99 (TheSession) and 0.97 (Meertens),
  indicating that once two melodies are similar enough to be flagged as candidates,
  their pitch identity score reliably distinguishes same-family from different-family pairs.
  Together these results show that pitch alignment is a useful but
  incomplete tool: precise when it finds a match, but blind to pairs that have
  diverged too far.

\nsb{Sequence-alignment algorithms cannot correctly align melodies.}
  \newline\citet{savageSequence2022a} had to manually correct their \num{328} alignments
  by visual inspection of the original scores to account for rhythmic information.
  We use their alignments to illustrate why this was necessary, identifying 
  two fundamental problems with pitch-only alignment.

  The first problem is ambiguity. When one sequence has more repeated notes than the other,
  the algorithm must insert a gap somewhere in the run -- but without considering rhythm, all insert positions in the expansion are equally valid.
  In the example in \fref{fig:fig1}B there are five equally-scored locations for
  the gap in the repeated 'G' region. Rhythm immediately resolves which is correct -- 
  the gap should be on the second 'G'.
  To assess the extent of this problem, we try to recreate the human-corrected alignments in
  \citep{savageSequence2022a} using global alignments and many scoring functions.
  The best-performing scoring function managed to find the correct alignment for \SI{74}{\%} of pairs.
  However, of these, only \SI{13}{\%} had a single top-scoring alignment, where the ground-truth
  correct alignment is recovered unambiguously -- these pairs only had 1 or 2 gaps, so this appears to work only for very similar tunes.
  The rest had a median of 9 top-scoring alignments, with some exceeding a thousand candidates.
  This places a clear limit on accuracy and usefulness, even for highly similar tunes.

  The second problem is more pernicious: some correct alignments are unreachable by any pitch-only algorithm.
  In \fref{fig:fig1}B, ``CEF'' in the top melody corresponds to ``CEFF'' in the bottom melody.
  A quarter-note `C' in the top sequence corresponds to two eighth notes `C' and `E' in the bottom -- 
  the note subdivision leads to a gap. The correct answer, given the rhythm, is to insert the gap after the `C'
  to align the pitch-mismatched `E' and `F': ``C-EF''. A pitch-only alignment will \textit{always}
  align the two `E' notes and place the gap on the `F': ``CE-F''.
  This is why for \SI{17}{\%} of pairs, the ground-truth alignment did not achieve
  a top score, regardless of the scoring function used.
  Generally, alignment on pitches alone will lead to systematic
  undercounting of pitch substitutions.

\begin{figure*}[th!]
\centering
\includegraphics[width=0.95\textwidth]{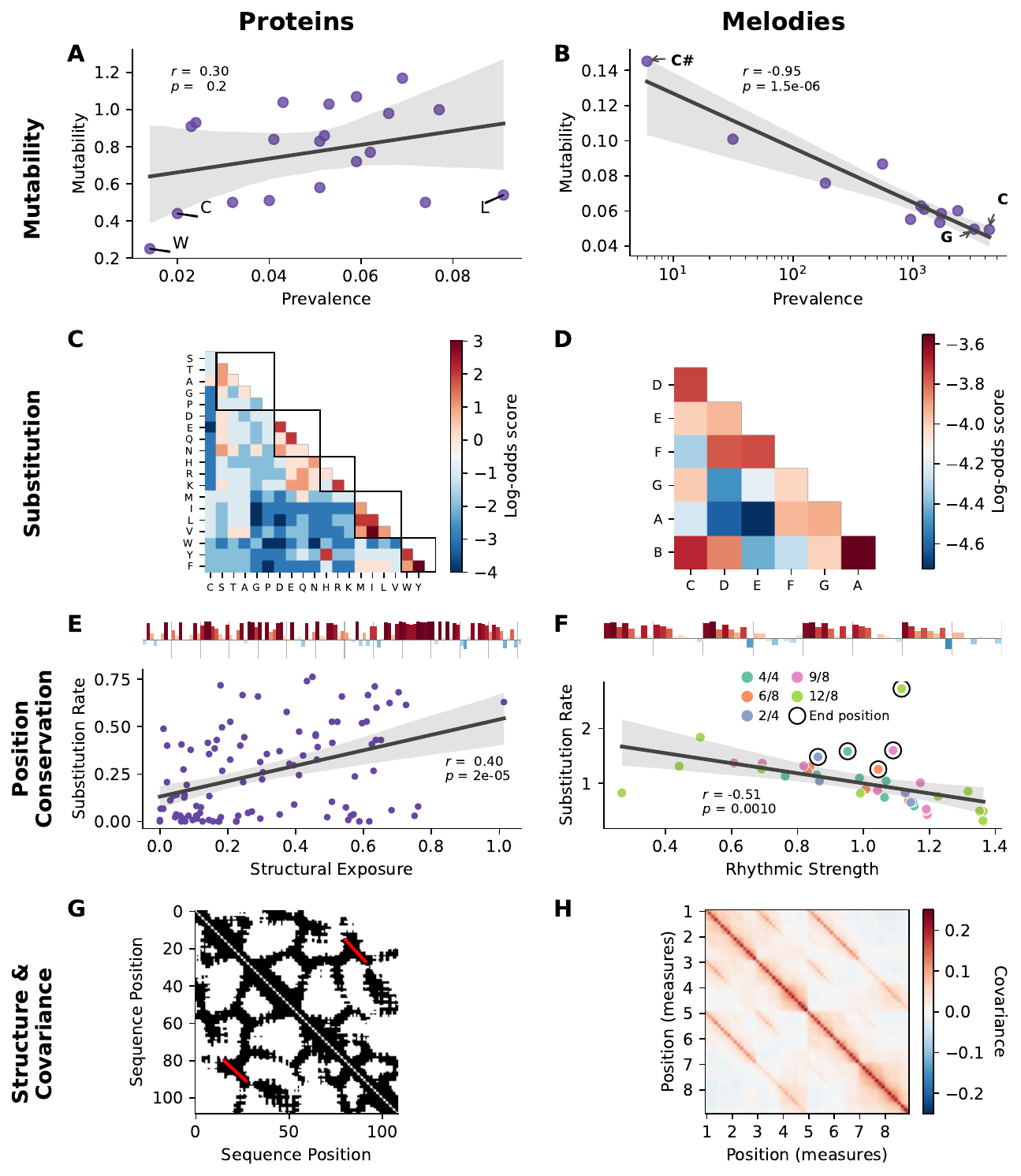}
\caption{\label{fig:fig2}
  \textbf{Comparing molecular and melodic evolution.}
  \textbf{(A)}~Amino acid mutability (relative to alanine) and
  prevalence;\cite{jonesRapid1992} C~(cysteine), W~(tryptophan) and L~(leucine)
  are highlighted.
  \textbf{(B)}~Pitch mutability and prevalence ($\pid \geq \SI{85}{\%}$)
  in Irish traditional music; C~(tonic), G~(dominant) and C\# are highlighted.
  \textbf{(C--D)}~Substitution matrices (lower triangle; log-odds of
  observed to expected substitution rate) for amino acids
  (C;~BLOSUM62) and pitches (D;~Major scale, $\pid \geq \SI{85}{\%}$).
  Amino acids are ordered by chemical similarity according to the Dayhoff
  classification scheme (C;~black squares).
  \textbf{(E--F, top)}~Sequence conservation along cytochrome~c (P00004)
  and the melody ``Man of the House'' (first part); red indicates
  conserved positions, blue indicates variable positions.
  Vertical lines mark every eighth position.
  \textbf{(E, bottom)}~Per-position substitution rate versus structural
  exposure (relative solvent accessibility) for cytochrome~c. 
  \textbf{(F, bottom)}~Per-position substitution rate versus rhythmic
  strength (probability of a note onset at that metric position),
  both normalised by the per-meter mean, for five meters
  (4/4, 6/8, 2/4, 9/8, 12/8; $\pid \geq \SI{85}{\%}$).
  The final position of each measure is circled.
  \textbf{(G)}~Structure of thioredoxin (P0AA25) depicted as a C$\alpha$ contact
  map -- positions are considered in contact (black) if they are within
  \SI{0.8}{nm} in the folded structure. A parallel $\beta$-strand
  structural element is highlighted in red.
  \textbf{(H)}~Position-position covariance for 4/4 tunes
  ($\pid \geq \SI{50}{\%}$).
}
\end{figure*}

\nsb{Rhythm-aware alignment of tune parts.}
  The dance tunes in TheSession follow strict rules governing meter and structure,
  where tunes are composed of parts of either 8 or 16 measures with a regular meter. For example, the 
  jig ``The Lark in the Morning'' is a four part jig (6/8) composed of four parts of 8 measures each, that
  are each repeated once (\fref{fig:fig1}C; only two parts are shown). We separated
  tunes into their parts (see Dividing Tunes into Parts), and quantize the pitches onto
  a regular grid (\fref{fig:fig1}D; see Melodic Sequence Representations) which makes it easy
  to align using note and measure onsets. Our pipeline ultimately produced
  \num{51680} pairs of similar parts from \num{41050} tunes
  (see Rhythm-Aware Part Alignment).

\subsection*{Comparing molecular and melodic sequence evolution}

  With this alignment method in place, we now adapt four canonical bioinformatics
  analyses -- mutability, substitution matrices, positional conservation, and covariance --
  to compare the statistical signatures of molecular and melodic evolution.
%\nsb{Molecular and melodic evolution exhibit distinct statistical patterns.}
% Four canonical bioinformatics analyses -- mutability, substitution matrices,
% positional conservation, and covariance -- underpin much of what is known about
% protein evolution, and together form the historic backbone of the field.
  These analyses underpin much of what is known about protein evolution,
  and together form the historic backbone of the field.
  The protein panels in \fref{fig:fig2} are chosen as familiar illustrative
  examples of phenomena that have been established across thousands of protein
  families;\cite{dayhoff221978,henikoffAmino1992,echaveCauses2016,marksProtein2011}
  our focus is on whether analogous statistical signatures exist in melodies,
  and whether melodic evolution follows the same pattern.

% \textit{Mutability.~}%
\nsb{Mutability.}
  Amino acids vary at different rates. By aligning closely related protein
  sequences and counting the occurrences and substitutions
  across variants, \citet{dayhoff221978} obtained a measure of
  mutability -- the substitution rate divided by the occurrence rate. Differences in
  mutability reflect the functional importance of each amino acid and are shaped by
  biochemistry. For example, cysteine is not easily replaced as it is the only
  amino acid with a reactive sulphur, essential for both structure (disulphide bonds)
  and catalytic function.\cite{gilesMultiple2003}
  
  A side-by-side comparison of mutability against prevalence in the two
  domains reveals a sharp asymmetry (\fref{fig:fig2}A,B). For amino acids the
  correlation is weakly positive and non-significant ($r = 0.30$,
  $p = 0.2$). For pitches we see an exceptionally strong negative relationship
  ($r = -0.95$): more prevalent notes are less likely to mutate. This correlation is
  robust, holding when tune parts are analysed separately by mode
  (Supplementary Fig.~2). The tonic (C) and dominant (G) are the most stable,
  while the rarest notes are the most readily substituted; for tunes in Dorian mode,
  the most mutable note (minor 6th) is more than twice as mutable as the
  least mutable note (tonic).

  No simple correlation emerges for amino acids because the prevalence and
  mutability of each amino acid are governed by several independent factors.
  Tryptophan is rare because it is metabolically costly to synthesise;
  cysteine is rare because its reactive sulphur is potentially hazardous;
  leucine is common because it plays a structural role in protein cores.
  Each amino acid is affected by functional, structural, and
  biosynthetic constraints, and no single axis dominates. Pitches, by
  contrast, are not as multi-faceted, and their mutability is free to
  follow the single underlying dimension of note frequency in Hz.

% \textit{Substitution matrices.~}%
\nsb{Substitution matrices.}
  Next we ask which \textit{pairs} of elements are interchangeable.
  Substitution matrices count co-occurrence -- how often each pair of letters
  co-occurs at the same position across many aligned sequence pairs,
  normalised by each element's baseline frequency to give a log-odds
  score.\cite{henikoffAmino1992} Positive values indicate pairs that substitute
  more often than expected by chance, and negative values the opposite. The
  most widely used protein matrix, BLOSUM62, reveals clusters of chemically, and hence
  functionally, similar amino acids -- \eg, hydrophobic amino acids exchange with other hydrophobic
  amino acids (\fref{fig:fig2}C). We computed an analogous matrix for pitch sequences in Major-mode
  tune parts (\fref{fig:fig2}D, see Supplementary Fig.~3 for other modes).
  A similar analysis was performed by \citet{savageSequence2022a},
  but with important and consequential methodological differences (see Supplementary
  Section~5).

  For pitch as well as amino acids, elements cluster by similarity, but the clustering
  pattern is different. For pitch, the similarity metric that emerges is not chemistry
  but \textit{interval size}: the highest log-odds scores lie close to (or far from,
  since pitch class structure is circular) the diagonal, meaning that
  pitches preferentially substitute for their scalar near-neighbours. Both domains
  therefore produce substitution matrices that group elements by functional
  similarity; they differ only on what dimension defines ``similar''.

% \textit{Positional conservation.~}%
\nsb{Positional conservation.}
  Evolutionary rates also vary along the length of a sequence: some positions
  are far more conserved than others.
  In proteins, this is because amino acids at structurally or functionally
  critical positions (\eg, catalytic sites in enzymes) cannot be easily replaced.
  Structurally, amino acids buried in the folded core cannot tolerate
  substitutions that would disrupt packing, while surface-exposed positions are
  freer to change.\cite{echaveCauses2016} \fref{fig:fig2}E illustrates this
  with cytochrome~c: position-by-position conservation is shown on top, and
  per-position substitution rate against relative solvent accessibility -- 
  the fraction of an amino acid's surface that is exposed -- on the
  bottom.

  Melodic positions also vary systematically in conservation, but for a
  very different reason. \fref{fig:fig2}F shows conservation along the first
  part of ``Man of the House'' (top), and per-position substitution rate
  plotted against \textit{rhythmic strength} -- the empirical probability that
  a note onset falls on a given metric position -- pooled across five meters
  (bottom). Positions of high rhythmic strength are the most conserved; those
  at the end of each measure are the least.

  Whereas for proteins structure is spatial (3D), for melodies it is temporal --
  where a position sits in the metrical cycle.
  This distinction leaves a striking visual signature in the conservation
  strips at the top of \fref{fig:fig2}E,F. For cytochrome~c, conserved and
  variable positions are scattered along the sequence with no apparent
  pattern or periodicity, because the seqeuence order becomes decorrelated
  from spatial order for positions that are further apart. For ``Man of the House'' the
  strip is visibly regular, with conserved positions lining up at periodic
  intervals, because here the linear order of the sequence \textit{is} the
  structural order: the metrical cycle unfolds along the axis of the plot,
  and any regularity in that cycle shows up directly as regularity in
  conservation/variability.

% \textit{Covariance.~}%
\nsb{Covariance.}
  The fourth analysis looks not at individual positions but at
  \textit{correlated} substitutions between pairs of positions. In proteins,
  covariance is a window onto 3D structure: two sequence positions that
  are distant along the linear chain, but brought into contact by folding,
  co-evolve to preserve that contact. The signal is rich enough that sequence
  covariance, with appropriate statistical treatment, can recover a protein's
  3D contact map, and is now a central ingredient in protein structure
  prediction.\cite{marksProtein2011,jumna21}

  We display the two analyses in complementary forms. \fref{fig:fig2}G shows
  the C$_\alpha$ contact map of thioredoxin -- the physical structure that
  protein covariance analysis is used to reconstruct -- with a parallel
  $\beta$-strand highlighted as a line parallel to the diagonal. Melodies do
  not fold in three dimensions and have no analogous structural ground truth;
  we therefore show the position-position covariance matrix for 4/4 tune
  parts directly (\fref{fig:fig2}H), computed over the first eight measures.

  Both short- and long-range structure are present, and the long-range
  component is strikingly periodic, at integer multiples of one measure.
  This is visible as bands parallel to the diagonal, reminiscent of the $\beta$-strand signature
  in the protein covariance matrix. We conjecture that this periodicity arises from repetition itself: whenever a
  motif recurs within a tune, its repeated positions become evolutionarily
  correlated. Changes to the melody thus tend to be repeated when replayed later in the tune.

\begin{figure*}[h!]
\centering
\includegraphics[width=0.95\textwidth]{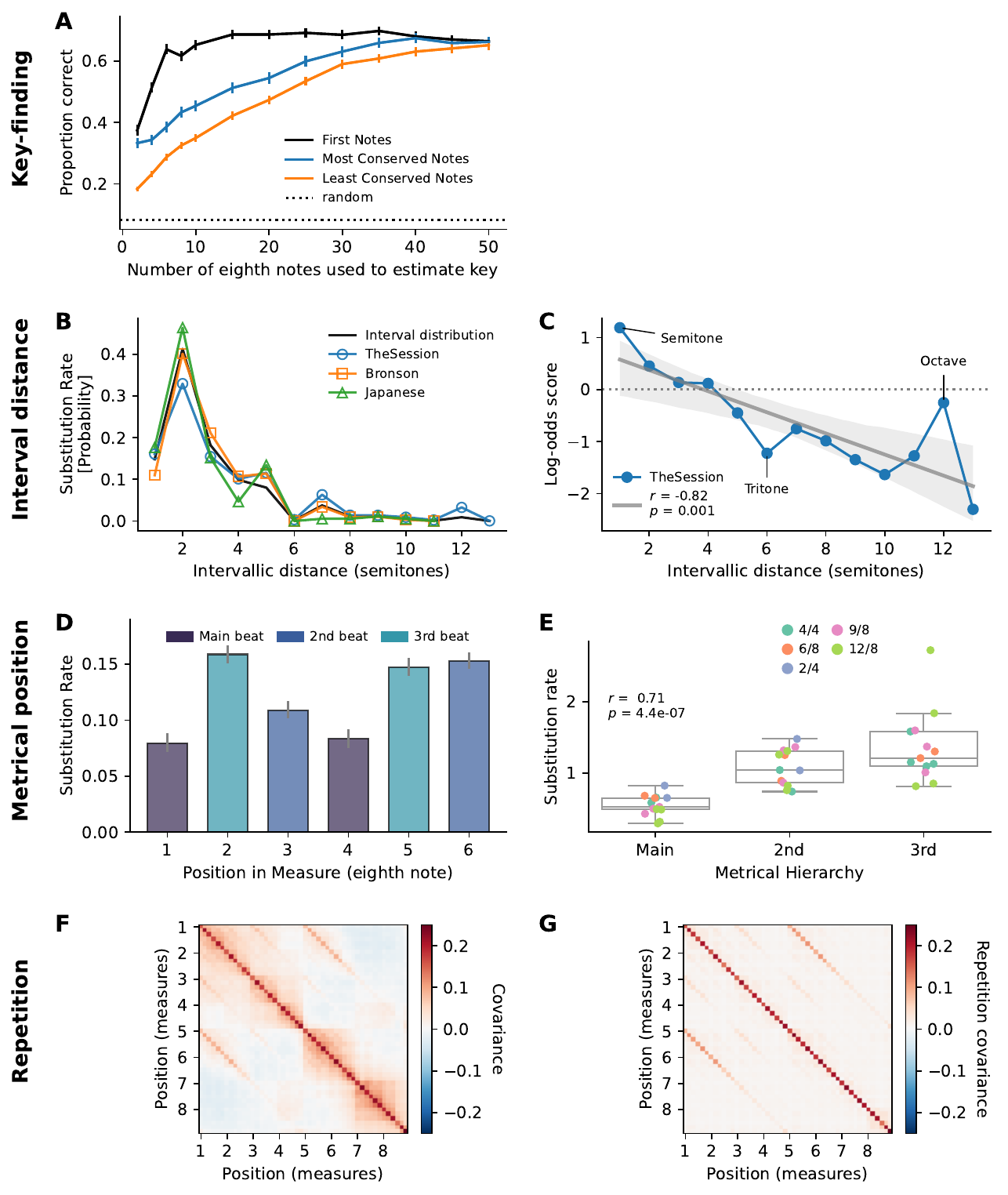}
\caption{\label{fig:fig3}
  \textbf{Exploring hypotheses for the rules of melodic evolution.}
  \textbf{(A)}~Key-finding accuracy using the first $N$ eighth notes selected
  by position in the melody (`First Notes'), or by within-family substitution
  rate (`Most Conserved Notes' and `Least Conserved Notes'). Evaluated on
  tune families with at least \num{10} alignment pairs.
  \textbf{(B)}~Absolute substitution rate versus intervallic distance
  (coloured lines) alongside the melodic interval probability distribution
  for TheSession (black line). All curves are normalised by their sum for
  ease of comparison. Bronson and Japanese data from
  \citep{savageSequence2022a}; TheSession rates for pairs with
  $\pid \geq \SI{85}{\%}$.
  \textbf{(C)}~Log-odds score (log-odds of observed to expected
  rate) versus intervallic distance ($\pid \geq \SI{85}{\%}$). Grey line and
  shaded region show a linear fit with \SI{95}{\%} confidence interval
  (Pearson's $r = -0.82$, $p = 0.001$, $n = 13$).
  \textbf{(D)}~Substitution rate at each eighth-note position within a 6/8
  measure, coloured by metrical hierarchy. Error bars show bootstrapped
  \SI{95}{\%} confidence intervals.
  \textbf{(E)}~Normalised substitution rate versus metrical hierarchy, pooled
  across five meters (4/4, 6/8, 2/4, 9/8, 12/8), coloured by meter
  ($r = 0.71$, $p = 4.8 \times 10^{-7}$).
  \textbf{(F)}~Position-position covariance matrix, $\Sigma_{ij}$, for all 6/8 tune pairs
  ($\pid \geq \SI{50}{\%}$).
  \textbf{(G)}~Repetition covariance for the same pairs: the component of
  $\Sigma_{ij}$ arising from position pairs linked by within-tune pitch
  repetition in both sequences.
}
\end{figure*}

\subsection*{Explanations for these patterns}

  Having shown that comparable analyses can be robustly performed on protein
  sequences and melodies, producing meaningful but different results, we now
  turn to testing exploratory hypotheses that attempt to explain these patterns.

\nsb{Conserved notes are more useful for key-finding.}
  The strong correlation between pitch mutability and prevalence (\fref{fig:fig2}B)
  begs an explanation. One candidate is key and mode identification: 
  prevalent notes may be those most useful for establishing a tune's tonality, which in turn
  constrains the set of likely pitches and thereby facilitates recognition,
  learning, and memory. This hypothesis predicts that the stable notes will be the most
  informative about tonality. For an exploratory test, we develop a key-finding algorithm
  trained on TheSession collection (see Key Estimation), and evaluate it
  by estimating the key from only the first $N$ notes of each tune
  (\fref{fig:fig3}A). The algorithm achieves about \SI{65}{\%} accuracy
  using only the first \num{10} notes. This is well above chance
  (\SI{12.5}{\%}), and it may be close to the practical limit, since tonality
  is not strictly established in many Irish tunes, which was historically
  practised without harmonic accompaniment.

  We then apply this algorithm to the $N$ most- and least-conserved note positions
  within tune families ($\pid \geq \SI{85}{\%}$). The most-conserved notes 
  are substantially more useful in key determination than the least-conserved notes (\fref{fig:fig3}A),
  with a difference in accuracy of \SI{34}{\%} at $N = 10$, which persists 
  even beyond $N = 50$. This shows that evolutionarily stable notes carry
  disproportionate information about the tonal identity of a tune. 

  Key-finding is functionally useful because knowledge of key and mode constrains
  the set of likely pitches, making melodies easier to predict and perceive.
  We can see this reflected in the data by looking at entropy, an information-theoretic
  measure of complexity\cite{mcbrideInformation2024}. Pitch class entropy across the full dataset is 3.05 bits,
  falling to 2.75 bits when tunes are considered within individual keys and modes -- a \SI{10}{\%}
  reduction. Stable notes thus appear to help achieve cognitive economy, and their stability
  may be maintained by selection pressure on learnability and memorability.

\nsb{Substitution distances are constrained by melodic motion.}
  The pitch substitution matrix (\fref{fig:fig2}D) showed clustering along
  the scalar diagonal, hinting that interval size is the relevant similarity
  metric for pitch substitution. One candidate explanation is that the
  motor constraints governing melodic motion also govern
  which notes substitute, so that substitutions are limited to those that
  maintain the melodic style. Unlike amino acids, pitches have a natural,
  one-dimensional similarity metric, so we can measure substitution rate
  as a continuous function of intervallic distance and test this directly. It also
  lets us probe whether octave equivalence, the perceptual tendency to hear pitches an
  octave apart as similar,\cite{marjiehComposite2023} plays a role.

  We first measure absolute substitution rate as a function of intervallic distance
  for TheSession tune parts ($\pid \geq \SI{85}{\%}$), since this is what was previously calculated for
  the Bronson and Japanese collections.\cite{savageSequence2022a}
  We divided each rate curve by its integral to aid visual comparison (\fref{fig:fig3}B).
  Here we can see similarities between the three collections. The most notable difference
  is that TheSession tunes have much higher note-substitution rates at large intervallic distances
  (7 and 12 semitones) than found previously. This perhaps reflects different constraints on melodic
  evolution in vocal (Bronson and Japanese) vs instrumental (TheSession)
  music. The instrumental tunes often contain melodic passages that follow
  arpeggiated (3, 4, 5 or 7 semitones) rather than scalar motion (1 or 2 semitones),
  presumably because the motor constraints are much weaker for instruments.
  Crucially, the substitution rate closely mirrors the overall melodic interval distribution
  (Pearson's $r = 0.98$, $p < 10^{-8} $, $n = 13$): substitutions tend to span
  the same intervals that are prevalent in the style. Arpeggiated passages are particularly
  prone to substitution, since the harmonically equivalent notes of a triad are
  all plausible alternatives; scalar passages in contrast typically move in a clear direction,
  limiting available options for variation.

  Normalising by the expected rate -- \ie, accounting for expectations due to scalar structure --
  (\fref{fig:fig3}C) makes the dominant trend stark:
  substitution rate decreases log-linearly with intervallic distance, falling by approximately \SI{20}{\%}
  per semitone. Two deviations stand out -- tritone substitutions (6 semitones) are rarer than expected;
  while octave substitutions (12 semitones) are much more common --
  consistent with spectral similarity and octave equivalence as alternate dimensions of pitch similarity.

\nsb{Metrical hierarchy generalises across meters.}
  The pooled relationship in \fref{fig:fig2}F uses rhythmic strength -- the
  empirical probability of a note onset at each metric position -- as a
  data-driven proxy for metrical hierarchy. We now highlight this relationship
  further using expert-annotated metrical strength -- an indicator of how
  strongly notes at metric positions are accentuated by performers.
  Within 6/8 tunes, substitution rates vary systematically with position, in a
  pattern that closely follows metrical hierarchy (\fref{fig:fig3}D; other
  meters in Supplementary Fig.~5). Notes on metrically strong positions are the
  most stable, and notes at the end of each measure the most mutable. When we
  pool across the five main meters and regress substitution rate against
  expert-annotated metrical hierarchy, the correlation remains strong
  (Pearson's $r = 0.71$, $p = 4.8 \times 10^{-7}$; \fref{fig:fig3}E),
  replicating and extending a similar finding in \citet{savageSequence2022a}.

  Why should metrical strength predict stability at all? Substituting a pitch
  on a strong beat would not disrupt the meter itself, so the pattern cannot
  be reduced to a rhythmic constraint. Instead, these results are consistent
  with the hypothesis that metrically
  strong positions collectively form a stable ``melodic
  skeleton''\cite{osuilleabhainCreative1990} that encodes tune identity and
  keeps variants mutually compatible in ensemble performance.

\nsb{Long-ranged covariance is due to repetition.}
  The short-range covariance visible in \fref{fig:fig2}H reflects the
  constraint on scalar motion identified above (\fref{fig:fig3}C):
  neighbouring pitches tend to move together to maintain melodic motion.
  For the long-range periodicities, we hypothesised that they reflect
  repetition within the tune: motifs that recur at fixed metrical
  distances will naturally generate periodic correlations between their positions.
  To test this, we compute a repetition-specific covariance matrix
  (see Covariance and Repetition) that isolates the component of covariance
  arising from within-tune pitch repetition.

 \fref{fig:fig3}F shows the full position-position covariance
  for 6/8 tunes (analogous to \fref{fig:fig2}H for 4/4 tunes), and
  \fref{fig:fig3}G the corresponding repetition component. The repetition-only
  matrix reproduces the periodicities of the full matrix, confirming that
  long-range covariance is consistent with the cyclical repetition structure
  of the tune. The need to preserve the structure and degree
  of repetition across variants therefore constrains how these melodies can evolve.

\section*{Discussion} 

  Proteins and melodies differ fundamentally because melodies have
  rhythm -- a hierarchical feature with no molecular counterpart. This difference
  hinders the transfer of bioinformatics methods to music,
  since standard alignment algorithms cannot appropriately represent
  the subdivision and fusion of notes. However, once we solved the alignment
  problem for highly-structured melodies, the rest of the
  standard bioinformatics toolkit transferred cleanly. Each of the four
  canonical analyses uncovers clear statistical regularities in
  both molecular and melodic evolution, but the patterns are
  categorically different between the two domains. Those
  differences are informative about the evolutionary forces that shape each:
  biochemical and biophysical constraints for proteins; social
  transmission and individual cognitive and motor biases for
  melodies. The combination of these forces makes melodies a
  promising model system for understanding the interplay of these cognitive factors in
  cultural evolution.

\nsb{Hypothesis: cognitive and motor biases shape melodic variation.}
  Two of our findings suggest independent transmission biases at the
  level of the individual learner. Evolutionary stability tracks tonal
  hierarchy: the tonic and dominant (`C' and `G' in the key of C) are
  the most stable pitch classes, while the least stable are those
  absent from specific modes (\fref{fig:fig2}B). This contrasts with
  amino acids, where prevalence and stability reflect a mix of
  functional, structural, and biosynthetic factors. A simple
  hypothesis is that stable notes are those most useful for
  establishing key and mode, since a clear sense of tonality reduces
  the information burden on performers and facilitates predictive coding, learning and
  memory. Our key-finding analysis supports this: the most
  evolutionarily stable notes are the most informative for key
  estimation, consistent with predictive cognitive constraints on cultural transmission (\fref{fig:fig3}A).
  The two components of this hypothesis -- stable notes are good for key-finding,
  knowledge of a key improves learning and memory -- can be directly
  tested with cognitive experiments. An alternate hypothesis is
  that both prevalence and stability are both products of a third
  component, such as a cultural attractor towards certain pitches.

  Motor constraints leave a separate signature in the substitution
  pattern. Substitution rates mirror the melodic interval distribution of each style
  (\fref{fig:fig3}B, $r=0.98$), meaning that substitutions tend to span the
  same intervals that are common in normal melodic motion. The higher rates
  of substitutions at large arpeggiated intervals (5 or 7 semitones) in TheSession
  compared to vocal collections likely reflect weaker motor constraints on
  large pitch jumps on instruments than for the voice --
  a fiddle or flute can leap more easily than a singer. 
  A direct test would be to compare substitution matrices across traditions differing
  systematically in instrumental versus vocal performance.
  The vocal-instrumental dichotomy can be broadened to include the fact that different
  instruments lead to different constraints -- single-string chordophones should
  exhibit more similar behaviour to the voice compared to multi-string chordophones,
  where large intervals are facilitated by strings tuned to different pitches.

\nsb{Hypothesis: a stable melodic skeleton encodes tune identity.}
  Pitch variation is organised along a temporal hierarchy,
  both at the within-measure metrical level (\fref{fig:fig3}E), and the
  between-measure level where repetition is clearly structured
  (\fref{fig:fig3}G). Metrical hierarchy itself has an obvious functional basis
  in dance -- strong beats are marked by higher onset probability,
  and this rhythmic regularity is what makes the beat perceptible and the tune danceable.
  But this explains only the rhythm, not the pitch; nothing about
  the metrical constraint requires that pitches on strong beats
  be more stable, since pitch could vary freely without disrupting
  the meter. The same puzzle arises for repetition. Reducing
  information load is a plausible general motivation for why
  repetition exists in music at all,\cite{mcbrideInformation2024}
  but it does not explain why one particular repetition pattern
  dominates, nor why some positions within a repeating pattern
  are more tightly conserved than others.

  Our results are consistent with the hypothesis that the hierarchical
  organisation of pitch stability reflects a heightened role of
  regular metrical positions -- a stable melodic skeleton -- in encoding tune identity.
  \cite{bayardProlegomena1950,bronsonBallad1969,cowderyMelodic1990,osuilleabhainCreative1990}
  In our data this skeleton is directly visible as the conserved (red)
  positions in the conservation strip for ``Man of the House''
  (\fref{fig:fig2}F, top), which recur at regular metrical intervals and are
  interleaved with the variable (blue) positions that carry most substitutions.
  Strong metrical positions are not only more frequently sounded but also
  accentuated by performers through dynamics, microtiming, and
  ornamentation, making them particularly salient -- more likely
  to be attended to, more faithfully perceived, and more reliably
  recalled.\cite{jonesDynamic1989} Their regularity also allows
  compression: rather than memorising the skeletal positions of
  every tune separately, a listener can rely on a template shared
  across all tunes in a given meter, onto which tune-specific
  pitches are mapped. A social force acts alongside this cognitive
  one: in ensemble playing, conformity bias is most consequential
  at positions where a pitch mismatch between players would be
  most audible -- the strong metrical positions. Repetition
  extends the same logic at a longer scale: in an 8-bar part,
  bars 1, 2, and 4 tend to be echoed in bars 5, 6, and 8, leaving
  bars 3 and 7 as the positions where pitch can vary most freely.
  This structure gives the melodic skeleton an internal shape that
  identifies the tune at the level of the part, while preserving
  space for the melodic variation prized in the Irish
  tradition.\cite{ohallmhurainShort2017}

  Four predictions follow from this hypothesis. First, the conformity mechanism
  predicts that positional stability should be higher in ensemble-based
  traditions than in predominantly solo repertoires. Second,
  iterated-learning experiments could be performed where note salience is manipulated
  -- whether due to metrical hierarchy or even non-metric accentuation -- and 
  should show that pitch is more stable on more salient positions\cite{jonesDynamic1989}
  after repeated cycles of cultural transmission.
  This should occur even when participants learn in isolation, disentangling 
  the cognitive mechanism from the social one.
  Third, hierarchical chunking of memory for these melodies should 
  coincide with the hierarchical nature of the metrical structure\cite{lorchChunking2022a,ahmadRhythmic2026}.
  Fourth, and most directly, enculturated listeners should be able
  to recognise a tune from its skeletal pitches alone, and
  artificial variants generated by perturbing skeletal versus
  non-skeletal pitches should differ in how natural they are judged
  to sound. Together these tests would distinguish the proposed
  mechanisms and assess whether the skeleton genuinely carries
  tune identity.
  
\nsb{Universal and stylistic features of melody evolution.}
  We primarily studied one tradition, but the analyses that do not require full alignment
  can be applied to any collection with tune family annotations. The mutability-prevalence
  relationship (correlation between evolutionary stability and tonal hierarchy) replicates in both the
  Bronson ($r = -0.81$, $p = 0.008$, $n=9$) and Japanese ($r = -0.84$, $p = 0.009$, $n=8$)
  collections, consistent with tonal hierarchies being a general feature of music.\cite{krumhanslTheory2010}
  The log-linear relationship between substitution rate and interval size is similarly
  likely to be widespread, since scalar motion dominates melodic motion across
  cultures.\cite{mcbrideInformation2024} By contrast, the specific covariance periodicities
  and mode-dependent substitution matrices are likely style-specific. Even these are
  likely drawn from a discrete inventory of possible forms, in line with evidence that metrical
  structures\cite{jacobyCommonality2024a} and pitch class sets\cite{mcbrideConvergent2023b,brownMusical2025a}
  cluster around a small number of recurring types across cultures. Charting where
  traditions converge and diverge across this formal inventory -- for repetition patterns
  as much as for modes or meters -- would itself be informative about the cultural
  and structural constraints that shape melodic evolution.

\nsb{Knowledge transfer: opportunities, pitfalls, and future directions.}
  Bioinformatics provided not only algorithms in this study, but also a useful conceptual template.
  The hierarchical structure of melodies, with parts nested within tunes,
  maps naturally onto the domain/gene distinction in molecular evolution.
  Parts can evolve semi-independently, be rearranged within a tune, or recombine
  to form new ones. This conceptual transfer from molecular to melodic sequence analysis was as valuable as any algorithmic one.

  Along with these advantages come pitfalls, of which rhythm is the central one:
  pitch-only alignment is not merely suboptimal but sometimes structurally incorrect, since a
  gap cannot represent the subdivision or fusion of a note.
  When a long note is divided into shorter notes,
  aligning some notes and assigning gaps to others fundamentally misrepresents
  the evolutionary relationship: subdivision of a note into multiple notes requires
  an alignment representation that permits one-to-many note mapping across sequences.
  \citet{savageSequence2022a} record such events as insertions/deletions, which they found
  to outnumber substitutions -- the opposite of coding DNA. Since our grid
  representation maps subdivided notes onto the metrical grid rather than onto gaps,
  they are absorbed as one-to-many correspondences and never counted as indels. This
  is appropriate for rigid-meter dance tunes, where subdivision preserves the meter;
  but in free-meter or solo singing traditions, lengthening a phrase to fit a new
  syllable genuinely inserts material, making the same event a true indel. The Bronson
  collection contains both, and distinguishing them may lead to different results
  than previously found.\cite{savageSequence2022a} 

  A second pitfall is fidelity of methodological transfer -- 
  knowing which elements of a canonical analysis to preserve, and
  which to adapt to the new domain. The substitution matrix illustrates both.
  The canonical Dayhoff formulation~\cite{dayhoff221978} normalizes observed substitutions
  by the rates expected under the background distribution, identifying
  substitution \emph{preferences} rather than raw counts.
  Without normalization, two common elements dominate the matrix
  simply because they are common, and the standard interpretation of
  functional similarity between commonly-substituted elements is no longer
  valid (Supplementary Section~5). A separate modification is also required:
  the canonical Dayhoff formulation assumes a single background distribution,
  which does not fit melodies obeying distinct modes, since tunes stick
  to a single set of notes defined by mode, limiting possible substitutions.
  For example, pooling across modes leads to high expected rates of substitutions
  between major and minor thirds, but in practice this almost never occurs.
  We avoided this issue by normalizing substitution
  matrices separately for each mode. Avoiding potential pitfalls of this kind -- recognising which
  analytic conventions can be preserved and which require modification --
  requires genuine expertise in both domains.  

  The most pressing unresolved problem is achieving general-purpose melodic alignment.
  The present analysis was possible because Irish dance tunes have unusually
  strict structure. For traditions with more variable form -- such as the Bronson ballads,
  where variants can differ in phrase length, meter, tempo, and mode -- no reliable
  automated alignment method yet exists. Incorporating rhythm into a dynamic programming
  framework is an obvious future direction, though the combinatorial explosion of possible
  alignments may require solutions analogous to the combinatorial extension technique
  developed for protein structural alignment.\cite{shindyalovProtein1998}
  One can learn from earlier work that considered the types of possible
  transformations that occur in melodies.
  \cite{sharpEnglish1907,bayardProlegomena1950,cowderyMelodic1990,janMemetics2007,mongeauComparison1990}
  The ground-truth alignments we release here provide a benchmark for evaluating
  whatever approaches are developed. The more fundamental open question is evaluation:
  what does it even mean for a melodic alignment to be correct when the sequences
  differ in meter or phrase structure? Answering this will require collaboration
  between researchers with algorithmic and musicological expertise.

  The data are not the limiting factor: for Irish music alone,
  public repositories, historical recordings, and archival
  notation amount to hundreds of thousands of tunes,\cite{eganSearch2023,harkinCreating2022}
  and digitisation efforts are underway for many other traditions.
  \cite{eerolaSuomen2004,van19a,malinCommunity2022,nishikawaCultural2022a,borsanIntroducing2025}
  Once algorithmic bottlenecks are resolved, the scale of available material makes
  this a potentially transformative moment for the computational study of musical evolution.

\nsb{Bridging corpus-based and experimental cultural evolution.}
  Research on cultural evolution divides into two complementary traditions.
  Controlled experiments on memory, perception, and iterated learning isolate
  roles of social transmission and cognitive biases -- showing, for instance, how rapidly humans acquire
  statistical tonal structure,\cite{loump10a} and disentangling the contributions
  of motor constraints,\cite{mitonMotor2020a}
  instrumental\cite{lumacaCultural2017,verhoefMelodic2021a} versus
  vocal\cite{anglada-tortLargescale2023a,popescuCore2024a} production, and
  differing modes of social transmission\cite{marjiehCharacterizing2025}.
  But such experiments are inherently limited: they cover only a small fraction
  of the stimulus space, their stimuli lack ecological validity, their participant samples
  are small and potentially unrepresentative, and evolution in the lab spans at most a few
  generations. Observational studies of cultural corpora make the opposite
  tradeoff: they register the joint outcome of all biases acting simultaneously
  over many generations of real transmission, but lack the causal explanatory power
  that comes with experimental control. Furthermore, corpus-based data-driven methods can be tackled by
  computational approaches that are easily scaled and adapted
  to other cultures. Pairing observation with modelling\cite{youngbloodConformity2019a,mcbrideMelody2024} 
  will let the two halves inform one another.

  We suggest that music is an unusually clean domain for studying the forces
  involved in cultural evolution\cite{boydCulture1988,henrichModeling2002,kirbyIterated2014,claidiereHow2014}.
  Cultural traits vary in how tightly functional pressures pin them down: language is bound to
  preserve meaning and grammatical structure, which sharply limits how it can
  change. Melodies do not have such strong functional constraints,
  which we argue means that music offers a much sharper lens into 
  the transmission and cognitive biases that shape cultural evolution.\cite{hoescheleCultural2022} 
  
% To give a concrete example of where this can go, consider the melodic skeleton
% hypothesis -- we can pair the observation/hypothesis with studies on memory.
% We can even directly link the two approaches by using a tool like IDYoM --
% a statistical method for predicting human expectation, that is learned from
% a corpus of melodies -- to make direct predictions about specific melodies
% given knowledge of a corpus -- this allows to actually design stimuli to
% test hypotheses. We can use study the hierarchical chunking of melodies
% in memory\cite{lorchChunking2022,ahmadRhythmic2026}.
% This approach can be integrated with general models of
% memory\cite{NEURIPS2022_ee5bb721}, offering a clean experimental approach
% allowing us to test models of memory.

\section*{Conclusion} 
  The large-scale analysis of biological sequences has transformed our understanding of molecular evolution
  -- revealing functional constraints, structural principles, and evolutionary histories
  that could not be inferred from individual sequences alone. Folk melodies share multiple key
  properties that make this same sequence analysis approach appealing and potentially powerful: they are one-dimensional sequences drawn
  from a limited alphabet, transmitted with copying error, and subject to selection.
  Yet previous attempts to apply bioinformatics methods to melodies have been limited
  in how they treated rhythm and by small sample sizes.

  Here we addressed these limitations systematically. A rhythm-aware alignment method,
  applied to \num{40000} Irish dance tune variants, enabled the first large-scale automated melodic alignment,
  and our analysis yielded four statistical signatures of melodic evolution.
  These signatures point to distinct evolutionary forces -- cognitive, motor, and social
  -- that jointly shape how melodies change, and motivate the hypothesis that a stable
  ``melodic skeleton'' encodes tune identity.
  The ground-truth alignments we release provide a benchmark for future
  algorithmic development toward more general melodic alignment.

  The broader significance lies in what this approach promises.
  Melodies are among the best-documented examples of cultural transmission,
  and the methods demonstrated here can be extended to other
  musical traditions. Understanding how
  cultural sequences evolve -- what is conserved, what changes, and why --
  is a fundamental question in human cultural evolution,
  and one that large-scale computational analyses can now begin to answer.

\small
\section*{Methods} 

\subsection*{Datasets}
  We study four datasets comprising Irish (TheSession),\cite{keithTheSessiondata}
  Dutch (Meertens),\cite{van19a} British/American (Bronson) and Japanese music.
  \cite{bronsonBallad1969,machidaRiBenMinYaoDaGuan1944,savageSequence2022a}
  TheSession is the main dataset which we use in all analyses.
  All except the Japanese corpus include tune family annotations that can be used
  as ground-truth clusters of related tune variants.
  We use the Meertens dataset only for identifying similar tunes.
  We use the Bronson and Japanese tunes for note mutability and substitution analyses, and
  to replicate some results from a previous study.\cite{savageSequence2022a}

\smallskip
\nsi{TheSession}
  ``thesession.org'' is a website that hosts transcriptions of (mainly Irish) tunes
  which have been contributed by users over a period of about 30 years.\cite{keithTheSessiondata} In Irish folk music
  melodic variation is common both within and between performances. Hence, tunes exist in
  many forms and the website allows users to contribute multiple variants of the same tune.
  There are currently \num{54075} tunes grouped into \num{22702} tune families,
  of which \num{1874} have more than \num{5} variants.
  We here summarize the key metrical, structural and tonal characteristics of the data, 
  and we refer the reader to the Supplementary Section~1 for a more comprehensive description of the dataset.

  Meter: Tunes are classified by the dances associated with them, and each class is played
  in a specific meter. Meter is a specification of how many beats there are in a bar and 
  which should be accented -- \eg, 4/4 means four quarter notes, and 6/8 means six eighth notes.
  Here we primarily study five meter types:
  4/4 (reel, hornpipe), 6/8 (jig), 2/4 (polka, march), 9/8 (slip jig), and 12/8 (slide).
  Reels and jigs are by far the most common.

  Structure: These dance tunes have regular, multi-part structures.
  Parts are either 8 (reels, hornpipes, jigs, polkas) or 4 (slip jigs, slides) measures,
  and are often repeated once, yielding a simple `AABB' schema for a two-part tune.
  It is the view of the first author, based on experience with the musical tradition, that parts are more stable than tunes.
  A tune can vary considerably through variation in the number of parts or how they are ordered.
  Parts, in comparison, usually differ by a few notes while retaining a similar overall
  metrical and melodic structure. Variant parts are also sometimes found in more than one distinct tune, as tunes
  have been created by recombination of parts from other tunes.
  Parts can thus be considered the more stable evolutionary unit than tunes,
  potentially analogous to the distinction between protein domains and full proteins.
  For this reason we divide tunes into parts and focus
  our analysis on alignment and comparison of parts.

  Tonality: Melodies are restricted to a set of pitches defined by key and mode.
  The key determines the pitch of the first note in a mode (the tonic), and the mode
  determines the pitches of notes based on their intervallic distance from the tonic.
  There are four 7-note modes associated with Irish dance tunes: 
  Major, Mixolydian, Minor, and Dorian. Tunes can also contain extra notes that are not in these modes,
  or fewer notes such that they use 6-note or 5-note scales.
  The key and mode are annotated in the dataset, but sometimes these annotations are incorrect.
  We use an algorithm to detect and re-annotate tunes with the correct modes, in cases where 
  we can unambiguously determine a mistake and correct it (Section \textit{Mode Estimation}).

\smallskip
\nsi{Meertens}
  The Meertens Tune Collection consists of \num{18618} digitally encoded pieces. These
  include both instrumental and vocal melodies. The data is available in multiple formats,
  and we used the Kern format.\cite{huronHumdrum1997}
  Tunes are classed into tune families, 83 of which have more
  than 2 labelled variants. This data is only analysed when testing
  if sequence alignment tools can be used for tune family identification.

\smallskip
\nsi{Bronson and Japanese}
  These are two collections of melodies that were digitized earlier. \cite{savageSequence2022a}
  One is the Bronson collection of ballads from Britain and North America,\cite{bronsonBallad1969}
  and the other is from an anthology of Japanese folk tunes.\cite{machidaRiBenMinYaoDaGuan1944}
  Unfortunately the melodies were digitized in a custom format that does not include any
  rhythm information, and pitch is represented using pitch class, which confines pitch information
  to a single octave. The authors did manually incorporate some of this information for their
  analyses, but it is not available in the data that was released.

  A subset of \num{242} highly-related ($\pid \geq 85\%$) pairs of tunes were manually aligned
  by \citet{savageSequence2022a}.
  We use these in three ways. First we use them as ground truth alignments to demonstrate the
  fundamental limitations of alignment algorithms that do not use rhythm information.
  Second, we recreate the note substitution matrix presented by \citet{savageSequence2022a};
  the authors only calculated raw substitution counts, whereas we instead calculate
  substitution preferences through the canonical Dayhoff normalization procedure.
  Third, we calculate note mutability and its correlation with note prevalence
  (the authors provided a similar analysis in an earlier preprint version).

  We show absolute substitution rate as a function of substitution distance for both the Bronson
  and the Japanese collections, as a comparison with TheSession. These values were taken directly
  from the figures in \citep{savageSequence2022a}. We additionally perform some analyses that
  are possible using the available manually-corrected alignments, and we do not attempt to
  analyse alignments of these melodies that were generated algorithmically.

\subsection*{Data Processing}
\nsb{Pre-processing and exclusion criteria.}
  Tunes are systematically excluded on the basis of complexity, regularity, and whether
  they are successfully parsed by two parsers (pyabc, music21; Section \textit{Parsers}). We process and exclude tunes sequentially as follows
  (and we report the numbers of tunes excluded at each stage):
  (i) Remove unnecessary elements (e.g., fermata, slur lines) (0 excluded).
  (ii) Parse tune with pyabc (\num{1972} did not parse).
  (iii) Exclude all (\num{4642}) tunes that have grace notes, multiple simultaneous
  notes , or multiple voices (staves). Grace notes are difficult to deal
  with since they have an indeterminate duration, and music21 has problems
  parsing grace notes in ABC format. Simultaneous notes and multiple voices render the melody ambiguous.
  (iv) Optionally exclude (\num{6001}) tunes with repeat lines that do not have a matching
  start and end pair, in case these are used by music21 to unravel the repeats as
  they are actually performed.
  (v) Parse the tune body using music21, unravel repeat lines, and transpose the tune
  % LAST TWO NUMBERS TO UPDATE!
  to a uniform key of C (\num{590} did not parse correctly).
  This left \num{41050} tunes for the main analyses.

  For the first analysis in this paper where we simply identify similar tunes,
  we run steps up to and including (iii), which left \num{47641} tunes. 

\nsb{Parsers.}
  The datasets have melodies stored in ABC\cite{walpi14} and Kern\cite{huronHumdrum1997} formats.
  We use two python packages to parse the data, music21\cite{cuthbertMusic212010} and pyabc.\cite{campagnolaPyabc}
  Music21 is able to read many symbolic notation formats, however it is less-well developed for ABC than for other file types.
  For example, music21 incorrectly parses pauses (fermata) by assuming they are notes with pitch C.
  Pyabc was developed specifically for parsing ABC so one might expect it to have fewer errors,
  however it is still under development (listed as pre-alpha on GitHub). We implemented several
  fixes and feature additions to pyabc, which have been integrated in an updated version of the software.
  To ensure quality control, we parse tunes using both parsers and check that they agree.
  We mainly use music21 because it has additional functionality over pyabc -- one can ``expand repeat lines'' so that
  the melodic sequence is parsed as it should be played.

\nsb{Melodic Sequence Representations.}
  We represent melodies in two complementary forms: (i) a
  \textit{sequence} of MIDI pitches paired with note durations, and (ii) a grid-quantised
  \textit{pitch vector} -- akin to piano rolls on digital audio workstations. 
  After parsing an ABC string, a melody of $\Nnotes$
  notes is represented as two sequences of equal length:
  a pitch sequence $\seqMidi = (p_1, \ldots, p_{\Nnotes})$
  of MIDI integer values, and a duration sequence $\seqDur = (d_1, \ldots, d_{\Nnotes})$
  where each $d_i$ is expressed in units of quarter notes.
  Transposition to a common key is performed by subtracting the tonic pitch class $\tau$
  (in semitones) from each element:
  \be \seqTmidi = \seqMidi - \tau ~.
  \ee
  For example,
  when transposing from the key of `A' to the key of `C', $\tau = 9$ semitones.
  MIDI and transposed MIDI sequences can be converted to pitch class sequences:
  \be
  \seqChroma = \seqMidi \mod 12 ~;
  \ee
  \be
  \seqTchroma = \seqTmidi \mod 12 ~.
  \ee

  To facilitate alignment, we convert a pair $(\seqMidi, \seqDur)$
  into a pitch vector $\mathbf{x}$ of length $L$ by quantizing onto a uniform time grid.
  The grid spacing $G$ is the smallest common denominator of all duration values in
  $\seqDur$ (or, when aligning two melodies, of the combined set of duration values of both tunes).
  Each note is expanded to $n_i = d_i / G$ consecutive grid cells carrying its pitch.
  The total length of the vector is $L = \sum_{i=1}^{\Nnotes} n_i$.
  This construction can be applied to any of the four pitch sequence representations
  ($\seqMidi$, $\seqTmidi$, $\seqChroma$, $\seqTchroma$),
  yielding pitch vectors $\mathbf{x}$, $\mathbf{x}^\tau$, $\mathbf{x}^c$,
  or $\mathbf{x}^{c,\tau}$ respectively.
  
\nsb{Dividing Tunes into Parts.}
  We assume that parts in TheSession tunes are either 8 or 16 measures long;
  see the Supplementary Section~1 for a detailed justification, and known exceptions.
  Tunes in 3/4 meter (annotated as `waltz', but actually a mix of dance and vocal songs) are excluded
  since they do not have the same rigid structure that makes this assumption valid.
  We first identify and remove any pickup (anacrusis) measures -- incomplete measures at the start of a tune
  -- by counting the total duration of notes in the first measure and
  comparing it to the expected duration $D_m$ for the given meter $m$.
  For example, $D_{4/4} = 4$ quarter notes.
  Any tune containing a measure whose total note duration differs from $D_m$ is excluded. 
  We then count the number of measures, $\nmeas$, and exclude any tunes if $\nmeas \mod 8 \neq 0$,
  removing any tunes that do not fit these assumptions.
  The tune is then divided into consecutive parts of 8 measures each. 

  Since 8-measure parts are sometimes repeated, we identify such cases and merge neighbouring
  parts into a single 16-measure part if they are highly similar.
  To identify similar parts, we convert each part into pitch vectors of equal length $L$,
  respectively $\vecMidi$ and $\vecMidi'$ (constructed as described in Melodic Sequence Representations).
  If at least \SI{80}{\%} of grid positions have the same pitch value in both parts
  ($\pid \geq$ \SI{80}{\%}) the two parts are merged into a single 16-measure part.

  We validated this algorithm using a ground-truth set of manually-annotated parts produced by the first author,
  who is expert in the Irish musical tradition. The set consists of 6 tunes (126 variants)
  with annotated parts. The similarity threshold was allowed to vary to determine the optimal value.
  A threshold of \SI{80}{\%} led to \SI{98}{\%} accuracy in identifying the correct number of parts.

\subsection*{Sequence Alignment}
\nsb{Pitch-based Sequence Comparison.}
  We use the bioinformatics software MMseqs2 to compare melodies to find similar tunes.\cite{steineggerMMseqs22017}
  This algorithm uses multiple pre-screening approaches to speed up comparisons,
  such that it only takes a few minutes to compare billions of tune pairs on a laptop.
  For tune pairs that pass screening, percent sequence identity ($\pid$) is calculated
  using the Smith-Waterman (local) alignment algorithm.\cite{smithIdentification1981}
  This algorithm requires sequences to be converted to a 21-letter alphabet corresponding
  to the 20 amino acids and 1 wildcard. To ensure that our pitch sequences have fewer than 21 
  unique values we convert from MIDI to pitch class, $\seqTchroma$. We transpose to a common key (C).
  We then map pitch classes onto the first 12 letters from the amino acid alphabet.
  For downstream analyses we can map these back to the original MIDI values.
  We write letter sequences to fasta files using Biopython.\cite{cockBiopython2009}
  Since MMseqs2 was optimized for biological sequences, 
  we optimized the score parameters (match, mismatch, gap open, gap extend) so that $\pid$ best
  predicts the grouping of tunes into tune families: match = 6, mismatch = -4, gap open = -4, gap extend = -3
  (see Supplementary Section~3).
  MMseqs2 is useful for finding similar sequences, but it was not designed for producing
  alignments for further analysis. For one specific analysis we require an alignment algorithm that can produce
  all top-scoring alignments in case there are more than one. For this we use Biopython's \textit{Align} module.

\nsb{Rhythm-Aware Part Alignment.}
  Due to regularities in the structure of TheSession tunes, it is quite easy to align
  parts once similar parts have been identified. From a total of \num{41050} tunes we extracted
  \num{51906} parts. An all-vs-all comparison ($\sim$\num{2} billion pairs) using MMseqs2 (see Pitch-based Sequence Alignment)
  identified $\sim$\num{4} million pairs of parts with similarity above a standard threshold (e-value $0.001$).
  We control for sequence redundancy by identifying clusters of identical parts 
  and removing all but one representative part.
  We exclude pairs of parts that differ in meter or number of measures.
  This leaves pairs of parts that can be converted to equal-length pitch vectors,
  $\vecMidi$ and $\vecMidi'$, which can be compared elementwise to calculate $\pid$.
  We do this for the remaining pairs at this stage, and exclude any
  pairs with $\pid \leq \SI{50}{\%}$. 
  This leaves us with \num{51680} pairs of parts for further analysis.

\nsb{Percent Identity (PID).}
  Percent identity ($\pid$) is the fraction of positions in an alignment at which
  two sequences share the same pitch value.\cite{raghavaQuantification2006}
  The denominator varies depending on the alignment method employed.
  When using the Needleman-Wunsch (global) algorithm,\cite{needlemanGeneral1970}
  $\pid$ is the number of matches divided by the length of the alignment.
  When using the Smith-Waterman (local) alignment algorithm,\cite{smithIdentification1981}
  $\pid$ is calculated in the same way as for a global alignment, but
  only over the aligned part of a sequence.
  \ie, local alignments can in principle align the full sequences, but in practice the ends
  of sequences are not included in the alignment if they are not sufficiently similar.
  When aligning pitch vectors, notes with long durations occupy multiple positions
  in the vector. $\pid$ is again the percentage of matches, but in this case matches are
  effectively weighted by the total duration of the matching notes.

\nsb{Sample Weighting.}
  Some tunes have many more variants than others, producing a long-tailed distribution
  over tune IDs. This problem is exacerbated when comparing pairs, since the joint distribution
  over pairs has an even heavier tail. To avoid results being dominated by the most common tunes,
  we apply inverse-frequency sample weights,
  $w_i = q_i^{-1}$, where $w_i$ is the weight of tune $i$,
  and $q_i$ is the fraction of all instances of tune $i$ in the full set of pairs. 
  This approach has the downside of increasing variance, since it increases the relative weight
  of rare IDs.\cite{liaoVariance2022} We can reduce this effect by modifying the weighting,
  $w_i = q^{-\alpha}_i$, where $0 \leq \alpha \leq 1$. When $\alpha=0$ samples are
  weighted (and biased) by the number of variants per tune. When $\alpha=1$ we recover the
  inverse-frequency weighting which increases variance due to over-weighting rare tunes.
  Since we are analysing pairs of parts that can come from different tunes,
  we weight by the geometric mean of the weights of the two parts.
  We use a value of $\alpha = 0.5$ as a compromise between bias and variance,
  and run sensitivity analyses to test robustness of results to weighting.

\subsection*{Bioinformatic Analyses on Melodies}
\nsb{Pitch Mutability.}
  For a set of aligned sequence pairs, we first convert them to transposed pitch
  class vectors $\vecTchroma$. We count pitch class occurrence -- how many
  grid positions each pitch class occupies overall -- and co-occurrence of
  all pitch-class pairs -- how many grid positions each pair of pitch classes co-occur
  across alignments. Substitutions are co-occurrences of one pitch class with another.
  Mutability of pitch class $c$ is the fraction of its occurrences that involve a substitution
  -- the number of substitutions divided by occurrences.
  For example, if D occurs 100 times and is substituted 20 times, $\mu(D) = 0.2$.

\nsb{Substitution Matrices.}
  The absolute substitution rate between two pitch classes is simply
  their co-occurrence count at substituted positions in $(\vecTchroma,\vecTchroma')$ pairs;
  this is what was calculated by \citep{savageSequence2022a}. However, this is
  dominated by the most common pitch classes. Following the original bioinformatics
  methodology from \citet{dayhoff221978}, we compute a normalized score:
  \be
  S(c,c')= \log \frac{\hat p(c,c')}{\hat p (c) \cdot \hat p (c')}
  \ee
  where $\hat p (c,c')$ is the observed probability of $c$ and $c'$ co-occurring at a position,
  and $\hat p(c)\cdot \hat p (c')$ is the probability expected if the two pitch classes
  occurred independently.
  For example, if $\hat p (C)  = 0.2$ and $\hat p (D)  = 0.2$, the expected probability of
  seeing C substituted for D is 0.02. 
  A positive score means the substitution occurs more often
  than expected by chance; a negative score means less often.
  Since the ordering in pairwise alignments (which sequence is `first')
  is arbitrary, substitutions are symmetric (C $\to$ D is the same as D $\to$ C).
  Thus there are $78$ unique scores -- 12 diagonal, and 66 off-diagonal. 

\nsb{Substitution Distance.}
  We compute substitution rate as a function of the absolute interval size
  between substituted notes. We use transposed MIDI pitch vectors, $\vecTmidi$, rather than
  pitch class vectors here, because pitch class differences obscure
  direction and changes greater than an octave -- \eg, a $\textrm{C} \to \textrm{D}$
  substitution could represent $+2$, $-10$, or even $+14$ semitones.
  For each aligned pair, we take the absolute difference between pitch vectors
  elementwise to get a vector of substitution distances,
  then compute the distribution over distances from \SIrange{1}{13}{semitones};
  there are larger intervals than this, but they become increasingly scarce with size.

  To normalize, we compute the expected distribution -- the distribution of interval
  sizes one would obtain by randomly pairing any two notes within the same melody.
  We use the observed distribution of all pairwise pitch differences within
  each melody as a proxy for this.  This expected distribution is computed
  only from melodies that appear in matched pairs, with each melody weighted
  by how many pairs it participates in. The normalized substitution rate at each
  distance is then the ratio of the observed to the expected probability, reported on a log scale.

\nsb{Position Substitution Rate.}
  For each aligned pair of parts, we compute a binary vector indicating whether
  a substitution occurs at each grid position. For cases where the grid is finer
  than an eighth note, we limit our search to grid positions
  that coincide with multiples of eighth notes. We calculate the number of substitutions
  that occur at each within-measure position $k$, given in integer units of eighth notes,
  and divide by the number of measures $\nmeas$ to get the position substitution rate for
  a pair of parts. The overall rate is then a weighted average across all pairs of parts.

\nsb{Covariance and Repetition.}
  For a set of $N_p$ aligned pairs of pitch vectors $(\vecTmidi, \vecTmidi')$
  of length $L$, we convert each pair into a boolean match vector $\mathbf{b}$,
  where $b_i = 1$ if the two vectors agree at position $i$ and $b_i = 0$ otherwise.
  We collect these vectors into an $N_p \times L$ match matrix $\mathbf{B}$,
  and let $\mu_i$ denote the weighted mean of column $i$,
  \ie, the mean match rate at position $i$.
  Throughout this section, angle brackets $\langle \cdot \rangle_w$ denote
  a weighted average over all $N_p$ pairs (see Sample Weighting).
  We then compute the $L \times L$ sequence covariance matrix
  \be
  \Sigma_{ij} = \operatorname{Cov}(\mathbf{B}_i, \mathbf{B}_j)
               = \langle (b_i - \mu_i)(b_j - \mu_j) \rangle_w.
  \ee
  Each entry $\Sigma_{ij}$ measures how often positions $i$ and $j$
  tend to match or mismatch together: a score of \num{1} indicates that they
  always covary (always matching or always mismatching across pairs),
  and a score of \num{-1} indicates the opposite.
  We deviate from traditional practice in one aspect: we measure covariance
  using grouped pairwise alignments for tunes (including unrelated tunes)
  with the same meter; for proteins covariance is
  computed using multiple-sequence alignments of related sequences.
  This methodological choice is apt because the dance tunes have
  such rigid structures which permits alignment of any tunes with
  the same meter, and is suitable for evaluating repetition
  within a musical style.

  To isolate the contribution of repetition to this covariance, we compute a
  second matrix that provides an exact partition of $\Sigma$.
  For each aligned pair, define the $L \times L$ boolean
  linked repetition matrix $R$, such that $R_{ij} = 1$ when
  positions $i$ and $j$ have the same pitch within $\vecTmidi$
  \textit{and} within $\vecTmidi'$ simultaneously.
  The repetition covariance matrix,
  \be
  \Sigma^{\mathrm{rep}}_{ij} = \langle R_{ij}(b_i - \mu_i)(b_j - \mu_j) \rangle_w,
  \ee
  together with its complement $\Sigma_{ij} - \Sigma^{\mathrm{rep}}_{ij}$
  exactly partitions the sequence covariance into contributions from
  repetition-linked position pairs and all others.

\subsection*{Protein Analyses}
  We use two proteins as illustrative examples: cytochrome~c (UniProt P00004) and
  thioredoxin (P0AA25). For cytochrome~c we retrieve
  orthologous sequences by running BLASTP\cite{altschulBasic1990} against SwissProt (minimum $\pid$
  \SI{50}{\%}, minimum query coverage \SI{80}{\%}; top \num{200} hits), fetch
  hit sequences from NCBI Entrez,\cite{sayersDatabase2025} and build a multiple sequence alignment
  with MAFFT.\cite{katohMAFFT2013} Structures are downloaded from the AlphaFold
  Protein Structure Database.\cite{varadiAlphaFold2022a}

\nsb{Substitution matrix.}
  We report BLOSUM62,\cite{henikoffAmino1992} the standard
  log-odds amino acid substitution matrix computed using blocks
  of conserved protein regions with $\pid \geq 62$.

\nsb{Positional conservation.}
  For each alignment column at which the query has no gap, we compute
  sequence identity as the frequency of the consensus (most common)
  amino acid among the non-gap entries (\fref{fig:fig2}E, top).

\nsb{Relative solvent accessibility.}
  Solvent accessibility is a common metric used to describe how much an amino
  acid is exposed to solvent, as opposed to buried inside the folded structure.
  Per-position solvent-accessible surface area is computed from each AlphaFold
  structure using the Shrake-Rupley algorithm\cite{shrakeEnvironment1973} and normalised by the empirical
  Gly-X-Gly maxima for each amino acid type to give relative solvent
  accessibility (RSA). Residue-level RSA values are mapped onto alignment
  columns, and for each reference position we take the mean RSA across all
  aligned orthologues with an available AlphaFold structure
  (\fref{fig:fig2}E, bottom).

\nsb{Contact map.}
  A contact map is a 2D representation of protein structure, indicating
  which sequence positions are close together.
  We compute the contact map from the AlphaFold structure
  of thioredoxin by taking all pairwise C$_\alpha$-C$_\alpha$ distances and
  thresholding at \SI{0.8}{nm}.

\subsection*{Musicological Analyses}
\nsb{Mode Estimation.}
  There are four 7-note modes associated with Irish dance tunes, which can
  be represented as sets of scale degrees -- pitch class measured in semitones from the tonic --
  that correspond to the number of semitones from the tonic:
  \begin{itemize}
    \item  Major: [0, 2, \textbf{4}, 5, 7, \textbf{9}, \textbf{11}]
    \item  Mixolydian: [0, 2, \textbf{4}, 5, 7, \textbf{9}, \textbf{10}]
    \item  Minor: [0, 2, \textbf{3}, 5, 7, \textbf{8}, \textbf{10}]
    \item  Dorian: [0, 2, \textbf{3}, 5, 7, \textbf{9}, \textbf{10}]
  \end{itemize}
  The highlighted scale degrees are those that distinguish the four modes.
  Since the mode annotations given by users are not always correct,
  we use a simple decision tree to automatically annotate modes based
  on pitch class histograms (see Supplementary Section~2).
  Some tunes do not have all seven pitch classes, and are labelled 
  `Major pentatonic', `Minor pentatonic', `Minor/Dorian', `Mixolydian/Dorian', or `indeterminate'.
  When analysing pairs of parts separated into the four main modes, we group pairs based on
  whether the modes of both parts are compatible with the target mode.
  The most conservative compatibility criterion is that the part mode must
  equal the target mode. The least conservative criterion is that the
  part mode must be a subset of the target mode -- \eg, if `Minor' is the target,
  then `Minor pentatonic' and `Minor/Dorian' are compatible.
  We use the least conservative criterion for our main analyses, and confirm
  that we get similar results using the most conservative criterion.

\nsb{Key Estimation.}
  There are many algorithms for estimating the key signature of a piece of music.
  Given that they are trained on popular and classical music, they are poorly suited
  to the modal character of Irish folk music. Thus we adapt the well-known
  Krumhansl-Schmuckler algorithm for estimating key,\cite{krumhanslCognitive2001}
  using the statistics of TheSession tunes.
  The Krumhansl-Schmuckler algorithm uses two modal profiles (Major and Minor),
  pitch-class histograms averaged over tunes with the same mode.
  We create four modal profiles by separating tunes into the four main modes, transposing
  them to the key of `C', and computing the pitch class histogram. For each mode,
  we get the average pitch class histogram over all tunes that are assigned
  the exact mode (see Mode Estimation). 
  We obtain an overall key-finding accuracy of \SI{76}{\%}; we could use a more
  sophisticed approach to achieve higher accuracy\cite{shahidEnsemble2023}, however
  this is sufficient for our purpose of demonstrating differences between
  conserved and non-conserved notes.

\nsb{Melodic Interval Distribution.}
  For each tune, we compute the melodic interval sequence by taking the absolute difference
  between consecutive pitches in the MIDI pitch sequence $\vecMidi$.
  We count the occurrences of each interval value within each tune
  and normalise by the total count to get a per-tune distribution.
  The overall distribution is then a weighted average across all tunes.

\nsb{Onset Probability (Rhythmic Strength).}
  For each tune, we compute a histogram of note onset positions within a measure,
  where position is measured in units of eighth notes from the start of the measure.
  We only consider integer values. The onset probability is the weighted mean of this
  histogram across all tunes, giving the empirical probability that a note onset
  occurs at within-measure position $k$.

\subsection*{Data and Code Availability}
  Code is available at \href{https://github.com/jomimc/TheSessionEvo}{github.com/jomimc/TheSessionEvo}
  and data is available at \href{https://zenodo.org/records/21356647}{zenodo.org/records/21356647}. 

\subsection*{AI Usage Declaration}
  Claude was used for analysis code and for copyediting the manuscript.
  The authors assume full responsibility for both.

\subsection*{Author Contributions}
  Conceptualization: JMM, WTF.
  Methodology: JMM.
  Software: JMM.
  Formal Analysis: JMM.
  Investigation: JMM.
  Resources: JMM, WTF.
  Supervision: WTF.
  Writing, original draft preparation: JMM.
  Writing, review and editing: JMM, WTF.

\subsection*{Acknowledgements}
  This research was supported by Austrian Science Fund (FWF)
  DK Grant ``Cognition \& Communication 2'' (W1262-B29) to WTF.

% Bibliography
\bibliography{TheSessionEvo}
\bibliographystyle{unsrtnat}

\end{document}

\title{\LARGE\bfseries\sffamily Supplementary Material for ``Contrasting statistical patterns in melodic and
 molecular evolution reveal distinctive constraints in a culturally evolving system.''}

\author{
    John M. McBride\textsuperscript{1,*} 
    \and 
    W. Tecumseh Fitch\textsuperscript{1,†}
}

\date{
    \small
    \textsuperscript{1}Department of Behavioral and Cognitive Biology, 
    University of Vienna, Vienna, Austria\\[0.5em]
    \textsuperscript{*}Correspondence: 
    \href{mailto:jmmcbride@protonmail.com}{jmmcbride@protonmail.com}\\
    \textsuperscript{†}Correspondence: 
    \href{mailto:tecumseh.fitch@univie.ac.at}{tecumseh.fitch@univie.ac.at}
}

\maketitle

\tableofcontents

\section{TheSession: Detailed Description}

\subsubsection{Tune Origins}

  Given migration and contact between cultural groups, the origins of tunes that come
  from an oral tradition are unknown except in cases where the composer is known. Despite that, 
  the vast majority of tune variants on TheSession are thought to be Irish, followed 
  by Scottish. Tunes from England, Wales, and regions of North America 
  settled by people from these traditions are also well represented. A 
  smaller number of tunes originate from diverse European countries 
  including France, Sweden, and Bulgaria. Tune origins are not 
  systematically annotated, since provenance is only known when the 
  composer is identified. A small number of tunes appear to have been 
  composed by users of the website, though this is discouraged by the 
  site's owner; such tunes typically do not accumulate multiple variants, and as such
  are not included in any evolutionary analyses.

\subsubsection{Tune Classes and Meter}

  Tunes are classified by the dances associated with them, each played 
  in a specific meter. The main classes are: reel (4/4), hornpipe (4/4), 
  jig (6/8), slip jig (9/8), and polka (2/4). Less common classes include 
  barndance (4/4), strathspey (4/4), march (2/4), slide (12/8), and 
  mazurka (3/4). Tunes in 3/4 are labelled waltzes, though many of 
  these are songs rather than dance tunes; tunes in 3/2 also appear 
  (e.g., Northumbrian hornpipes). Tunes are screened to ensure they 
  contain no meter changes within a tune. We only report analyses of the most
  common dance and meter types.

\subsubsection{Tune Structure and Parts as Evolutionary Units}

  Dance tunes have regular, multi-part structures. Parts are generally 
  8 measures long, and most often repeated once to a total of 16 measures.
  Slip jigs and slides are exceptions, as they are 
  4-measure parts intended to be repeated, yielding an effective length 
  of 8 measures. Parts are often repeated within a performance, with 
  minor variations at phrase endings depending on whether the part is 
  repeating or leading into a new part.

  Parts are more stable evolutionary units than whole tunes, which can 
  vary considerably in the number and ordering of their parts. The 
  following operations on parts are commonly observed:

  \begin{itemize}
      \item \textit{Repetition:} a part may be repeated or not 
      (AB $\to$ AABB; e.g., ``Drowsy Maggie'', tune ID 27)
      \item \textit{Reordering:} the positions of parts may be swapped 
      (AB $\to$ BA; e.g., ``The Wren's Nest'', tune ID 119)
      \item \textit{Insertion/deletion:} parts may be added or removed 
      (AB $\to$ ABC or AB $\to$ B; e.g., ``Drowsy Maggie'', tune ID 27)
      \item \textit{Recombination:} parts from different tunes may be 
      combined to form new tunes (e.g., ``The Dingle Regatta'', tune ID 23)
  \end{itemize}

  \noindent
  These are the same mathematical operations observed in biological 
  sequence evolution -- repeats, inversions, insertions, deletions, and 
  recombination -- making the analogy between tune parts and protein 
  domains particularly apt. Just as protein domains are semi-independent 
  evolutionary units that can be shuffled between proteins, tune parts 
  can evolve semi-independently and be recombined across tunes. For 
  this reason, all analyses in this paper are conducted at the level of 
  tune parts rather than whole tunes.

\subsubsection{Tonality and Key Constraints}

  There are four heptatonic modes associated with Irish dance tunes, 
  represented here as sets of semitone intervals from the tonic:

  \begin{itemize}
      \item Major: $[0, 2, 4, 5, 7, 9, 11]$
      \item Mixolydian: $[0, 2, 4, 5, 7, 9, 10]$
      \item Minor: $[0, 2, 3, 5, 7, 8, 10]$
      \item Dorian: $[0, 2, 3, 5, 7, 9, 10]$
  \end{itemize}

  \noindent
  The modes are distinguished by three scale degrees. Major and 
  mixolydian share a major third (4 semitones), while minor and dorian 
  share a minor third (3 semitones). Within the major-third pair, a 
  major seventh (11) indicates major and a minor seventh (10) indicates 
  mixolydian. Within the minor-third pair, a minor sixth (8) indicates 
  minor and a major sixth (9) indicates dorian.

  TheSession restricts key submissions to the natural keys A, B, C, D, 
  E, F, and G, and permits all four modes only for the keys A, D, E, 
  and G. This follows traditional practice, though in performance tunes 
  are commonly played in other keys. As a result, some tunes that are 
  genuinely in less common keys must be submitted under an 
  enharmonically equivalent annotation, which contributes to the 
  annotation errors described below.

  The key and mode annotations in TheSession contain three categories 
  of error. First, some tunes are mislabelled due to limited annotation 
  options — for example, the slip jig ``Farewell to Whalley Range'' 
  (tune ID 2410) is in F\# minor pentatonic, which has no direct annotation 
  option and is therefore often labelled with enharmonically equivalent 
  keys such as A major or B minor. Second, some tunes are assigned an 
  incorrect key that is then corrected through accidentals within bars 
  (e.g.\ ``The Clumsy Lover'', tune ID 6632). Third, different parts of a 
  tune can be in different keys or modes — usually related by the 
  relative minor or dominant and sharing the same note set 
  (e.g.\ ``Drowsy Maggie'', tune ID 27), but occasionally involving a 
  genuine modulation to a different pitch set 
  (e.g.\ ``The Humours of Ballyloughlin'', tune ID 210).

  The first issue would require an algorithm capable of estimating both 
  key and mode simultaneously — a non-trivial problem since many 
  key/mode combinations share identical pitch class sets, and existing 
  algorithms have not been tailored to modal Irish music. We do not 
  attempt to fully solve this here; its only consequence is that some true 
  positive tune pairs are missed during identification, without 
  introducing systematic artefacts, since miskeyed tunes will not align 
  well and will be excluded from analysis. The second and third issues 
  are addressed by re-estimating the mode given the annotated key, 
  which reduces the problem to classifying among four modes using 
  pitch class histograms. This simpler problem is solved by the decision 
  tree described in the following section.

  \clearpage

\section{Mode identification algorithm} 
  Given a tune where the key is known, we use the following decision tree
  to identify the associated mode based on the tune's pitch class histogram.

  \begin{enumerate}[leftmargin=*, label={}]

  \item \textbf{If major 3rds $>$ minor 3rds:}
  \begin{itemize}
      \item Major 7ths $>$ minor 7ths $\;\Rightarrow\;$ \textit{major}
      \item Minor 7ths $>$ major 7ths $\;\Rightarrow\;$ \textit{mixolydian}
      \item No 7ths present $\;\Rightarrow\;$ \textit{major pentatonic}
      \item Equal 7ths $\;\Rightarrow\;$ \textit{major/mixolydian} (ambiguous)
  \end{itemize}

  \item \textbf{If minor 3rds $>$ major 3rds:}
  \begin{itemize}
      \item Minor 6ths $>$ major 6ths $\;\Rightarrow\;$ \textit{minor}
      \item Major 6ths $>$ minor 6ths $\;\Rightarrow\;$ \textit{dorian}
      \item No 6ths present $\;\Rightarrow\;$ \textit{minor pentatonic}
      \item Equal 6ths $\;\Rightarrow\;$ \textit{minor/dorian} (ambiguous)
  \end{itemize}

  \item \textbf{If equal major and minor 3rds (or no 3rds):}
  \begin{itemize}
      \item Major 6ths and 7ths only $\;\Rightarrow\;$ \textit{major}
      \item Minor 6ths and 7ths only $\;\Rightarrow\;$ \textit{minor}
      \item Major 6ths and minor 7ths only $\;\Rightarrow\;$ 
            \textit{mixolydian/dorian} (ambiguous)
      \item Otherwise $\;\Rightarrow\;$ \textit{indeterminate}
  \end{itemize}

  \end{enumerate}

  \noindent
  Major pentatonic is treated as compatible with both major and mixolydian; 
  minor pentatonic is compatible with both minor and dorian. Applying this 
  algorithm to TheSession, 52\% of tunes are classified as major, and only 
  1.5\% are classified as indeterminate or ambiguous (minor/dorian or 
  mixolydian/dorian). Overall, 94\% of user-supplied mode annotations are 
  compatible with the mode assigned by the algorithm.

  \clearpage

\section{Optimizing scoring functions for pairwise alignment}
  Sequence alignment methods have four scoring parameters:
  \begin{itemize}
      \item \textbf{Match:} the score assigned to alignment of two 
      identical letters.
      \item \textbf{Mismatch:} the score assigned to alignment of two 
      different letters.
      \item \textbf{Gap open:} the score assigned to the first gap in a 
      consecutive run of gaps, or to a single isolated gap.
      \item \textbf{Gap extend:} the score assigned to each subsequent 
      gap in a consecutive run, beyond the first.
  \end{itemize}

  \noindent
  The default MMseqs2 parameters derive match and mismatch scores from 
  a standardised substitution matrix (e.g., BLOSUM62), in which scores 
  are assigned to each possible pair of letters based on their empirical 
  substitution likelihoods; gap open is $-11$ and gap extend is $-1$ by 
  default.

\begin{figure*}[h!]
\centering
\includegraphics[width=0.95\textwidth]{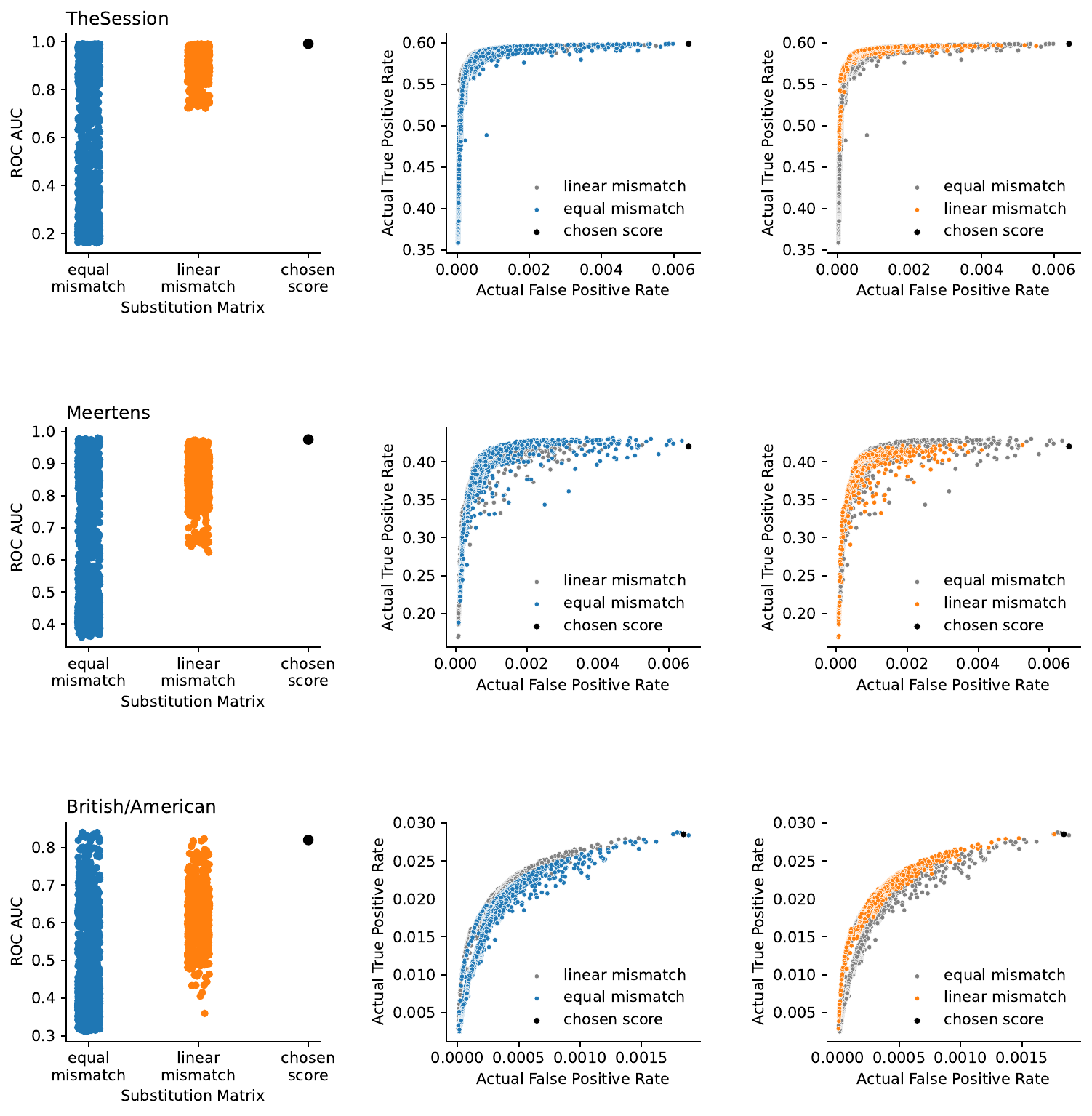}
\caption{\label{fig:1}
    \textbf{MMseqs2 parameter optimisation results across three datasets.}
    Left: Distribution of ROC AUC scores across all gap penalty combinations,
    separated by substitution matrix type (equal mismatch vs linear mismatch).
    Right: Actual true positive rate vs actual false positive rate for each
    parameter combination, showing the fraction of all within-family and
    cross-family pairs retrieved by MMseqs2 before any score threshold is
    applied.  Each point represents one combination of substitution matrix
    and gap penalties.  Colours indicate substitution matrix type.
}
\end{figure*}

  We optimised these parameters using a grid search over the following 
  ranges:

  \begin{itemize}
      \item match $\in \{2, 4, 6, 8, 10\}$
      \item mismatch $\in \{-4, -3, -2, -1, 0\}$
      \item gap open $\in \{-15, -14, \ldots, -2\}$
      \item gap extend $\in \{-14, -13, \ldots, -1\}$
  \end{itemize}

  \noindent
  with the constraint that gap open $\leq$ gap extend throughout. For 
  each parameter combination we computed a ROC curve and associated 
  metrics, including AUC and the true and false positive rates accounting 
  for the pre-filtering stage (see main text). Many parameter 
  combinations achieved similar performance (Supplementary \fref{fig:1}); 
  from these we selected: match $= 6$, mismatch $= -4$, gap open $= -4$, 
  gap extend $= -3$.

  Following the observation that log substitution rates decrease 
  log-linearly with intervallic distance (Fig.~3C of the main text), 
  we also tested whether mismatch scores proportional to substitution 
  distance could improve alignment performance. Using the same gap open 
  and gap extend scores as above, we performed a second grid search over 
  a reduced parameter range:

  \begin{itemize}
      \item match $\in \{1, 2, 3, 4, 5\}$
      \item per-semitone mismatch cost $\in \{-1.0, -0.8, -0.6, 
      -0.4, -0.2\}$
  \end{itemize}

  \noindent
  so that the mismatch score for a substitution of interval $\delta$ 
  semitones was match $- \delta \times$ cost. No improvement in 
  alignment accuracy was found relative to the uniform mismatch 
  baseline (Supplementary \fref{fig:1}), consistent with the 
  conclusion in the main text that pitch-only alignment is 
  fundamentally limited regardless of scoring scheme.

\subsection{Should substitution matrices be used for pitch sequence alignment?}
  Pitch substitution matrices differ substantially across modes
  (Pearson's $r \sim 0.6$--$0.9$ between modes), more so than biological
  substitution matrices differ across divergence levels
  (Pearson's $r > 0.9$), suggesting that mode-specific matrices should improve
  alignment (Supplementary \fref{fig:4}). Alternatively, since substitution
  rate depends strongly on interval size, a simpler distance-proportional mismatch score might
  suffice\cite{mongeauComparison1990,lavinSimilarity2010,hajicjrBuilding2023}.
  We tested both approaches across all three datasets and found no improvement in
  alignment accuracy (Supplementary Fig. 1). This is unexpected given how effective
  substitution matrix optimisation has been in bioinformatics, and suggests that
  pitch-only alignment is fundamentally limited -- rhythm must be incorporated for meaningful progress. 

  \clearpage

\section{Pitch prevalence and mutability for individual modes}
\begin{figure*}[h!]
\centering
\includegraphics[width=0.95\textwidth]{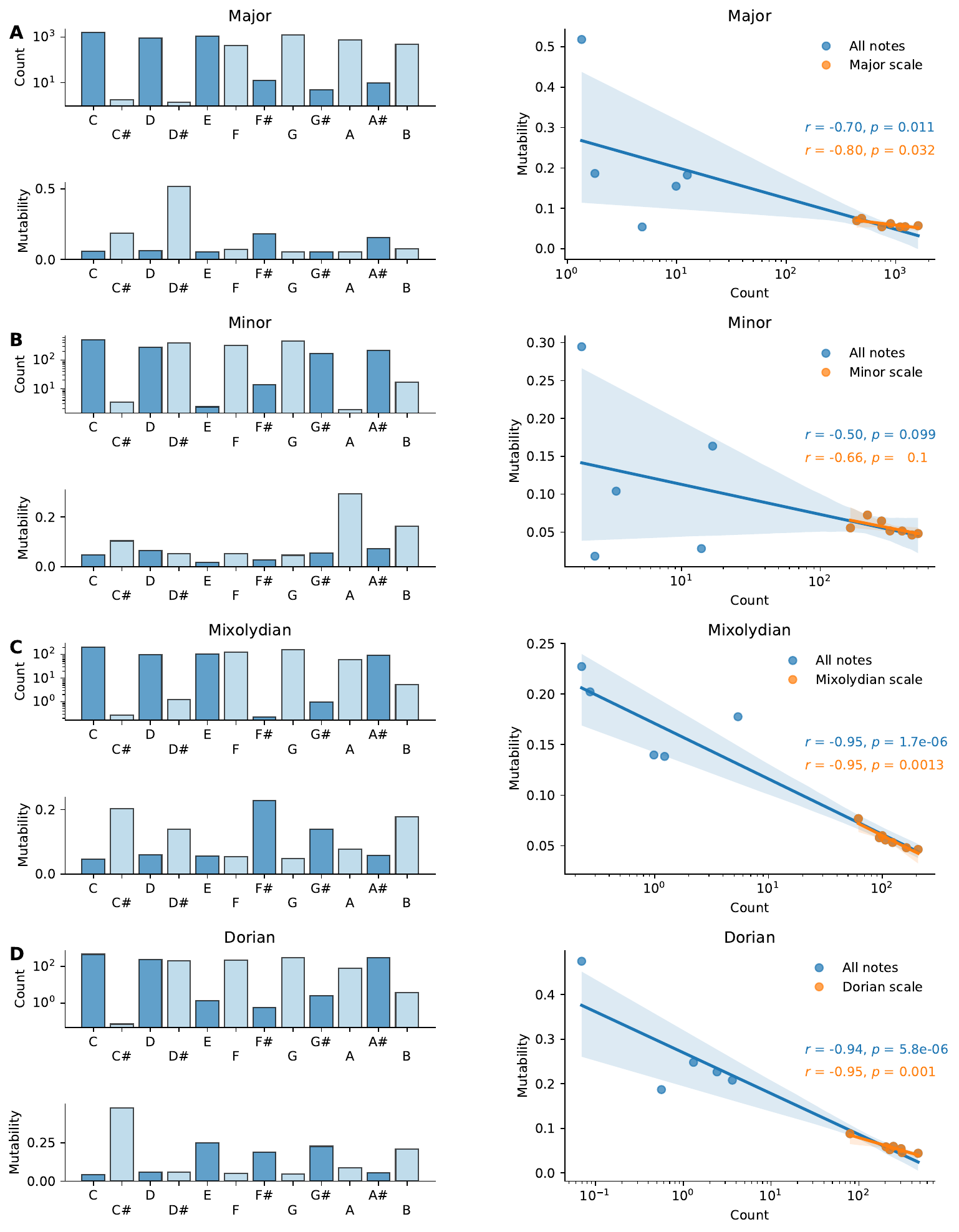}
\caption{\label{fig:2}
  A-D: Note prevalence (Count) and mutability for each mode separately,
  computed from pairs with $\pid \geq \SI{85}{\%}$.
  Left panels show the prevalence (top) and mutability (bottom) of each
  of the 12 chromatic pitch classes.
  Right panels show the inverse relationship between prevalence and
  mutability, with regression lines fit to all notes with non-zero count
  (blue) and to only the notes belonging to the mode's scale (orange);
  Pearson's $r$ and $p$ for each fit are shown in matching colours.
  Modes shown are major (A), minor (B), mixolydian (C), and dorian (D).
}
\end{figure*}

  The pitch mutability analysis in the main text pools tune parts across
  all four modes after transposing each part to a common tonic.
  Because the four modes share most scale degrees but differ in two or three,
  pooling conflates notes that are common in one mode but
  absent in another, which can obscure mode-specific patterns.
  To assess whether the inverse relationship between note prevalence and
  mutability holds within each mode individually, we repeated the analysis
  separately for major, minor, mixolydian, and dorian parts.
  For each mode we selected all pairs with $\pid \geq \SI{85}{\%}$ in which
  both parts were assigned to that mode (using the mode-identification
  algorithm described above), and computed note prevalence and mutability
  from the resulting substitution count matrix.
  Supplementary \fref{fig:2} shows that the inverse prevalence--mutability relationship
  holds in all but the minor mode; correlations are strong and significant whether
  computed over all 12 chromatic pitch classes or restricted to the seven
  notes of the mode's own scale, and they sharpen further in the
  within-scale case for all but the mixolydian mode.

  \clearpage

\section{Substitution matrices: normalization and mode separation}
\label{sec:submat}
\begin{figure*}[h!]
\centering
\includegraphics[width=0.95\textwidth]{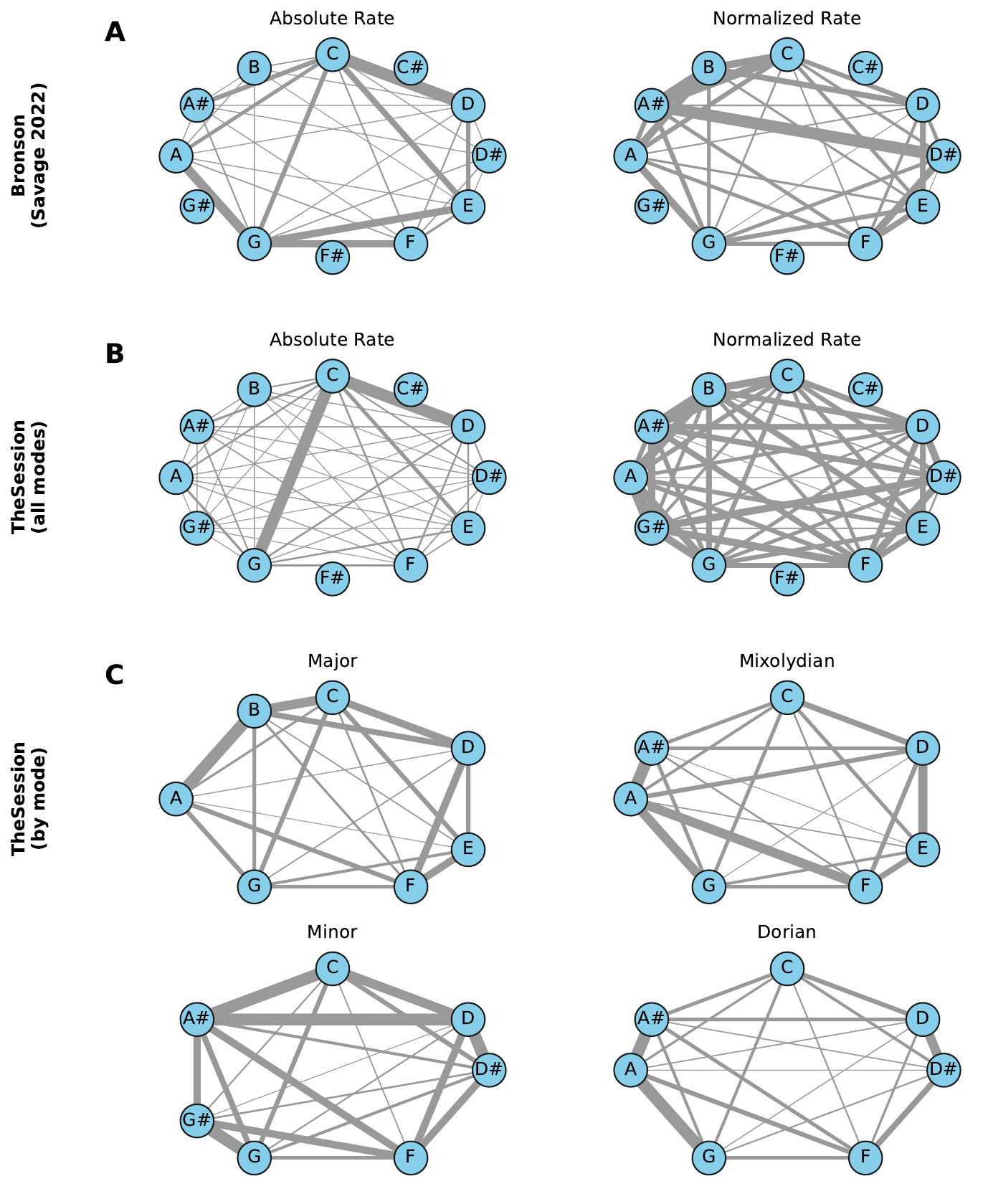}
\caption{\label{fig:3}
  \textbf{Pitch substitution graphs: the need for normalization and for
  mode separation.}
  Each panel shows a pitch substitution graph whose edges are weighted by
  substitution rate; edges are only drawn from nodes whose marginal
  frequency exceeds \SI{1}{\%}.
  \textbf{A}: Bronson British/American collection, absolute rate (left)
  and normalized log-odds rate (right). The absolute rate reproduces the
  analysis of \citet{savageSequence2022a}.
  \textbf{B}: TheSession tune parts pooled across all modes, absolute rate
  (left) and normalized log-odds rate (right).
  \textbf{C}: Normalized log-odds substitution graphs computed separately
  for each of the four main modes (major, minor, mixolydian, dorian).
  TheSession rates in B and C are calculated for all pairs with
  $\pid \geq \SI{85}{\%}$.
}
\end{figure*}

  A pitch substitution matrix records how often each pair of pitch classes
  co-occurs at substituted positions in aligned tune-pair variants.
  Substitution matrices have been computed for melodies before by
  \citet{savageSequence2022a} for the Bronson (British/American) collection,
  who reported \textit{absolute} substitution counts.
  This departs from the canonical Dayhoff--BLOSUM
  formulation,\cite{dayhoff221978,henikoffAmino1992} in which observed
  co-occurrence rates are normalized by the rates expected under each
  element's baseline frequency, yielding a log-odds score that isolates
  substitution \textit{preferences} from substitution \textit{opportunity}.
  Without this normalization the matrix is heavily influenced by element prevalence and
  scale structure: In the Bronson absolute-rate matrix a thick edge connects `C' and
  `G' (Supplementary \fref{fig:3}A, left), simply because these are the most common
  notes. The peak at two semitones in the substitution rate vs intervallic distance plot
  is a consequence of the fact that major seconds are the most common scale step.
  The corresponding log-odds matrix (Supplementary \fref{fig:3}A, right) isolates
  substitution preference.

  Applying the same normalization to TheSession data, pooled across modes,
  still produces a graph that is hard to interpret (Supplementary \fref{fig:3}B).
  This is because unlike proteins, which tend to contain at least one
  of each amino acid, melodies typically draw only from the notes of their mode.
  Computing expected co-occurrence rates from the pooled note distribution
  therefore conflates across modes -- erroneously suggesting, for instance,
  that major and minor thirds are interchangeable, when in practice they
  almost never substitute for one another since tunes rarely modulate
  between modes that would permit the exchange.

  The solution is to construct substitution matrices separately for each
  mode (Supplementary \fref{fig:3}C). The resulting graphs are considerably cleaner:
  most of the thick edges run along the perimeter of the circle,
  corresponding to small stepwise changes (scalar motion), and the
  differences between modes are substantial -- considerably larger than the
  differences found between biological substitution matrices constructed at
  different levels of sequence divergence (Supplementary \fref{fig:4}).

  \clearpage

  \clearpage

\section{Comparing biologial and musical substitution matrices}
\begin{figure*}[h!]
\centering
\includegraphics[width=0.95\textwidth]{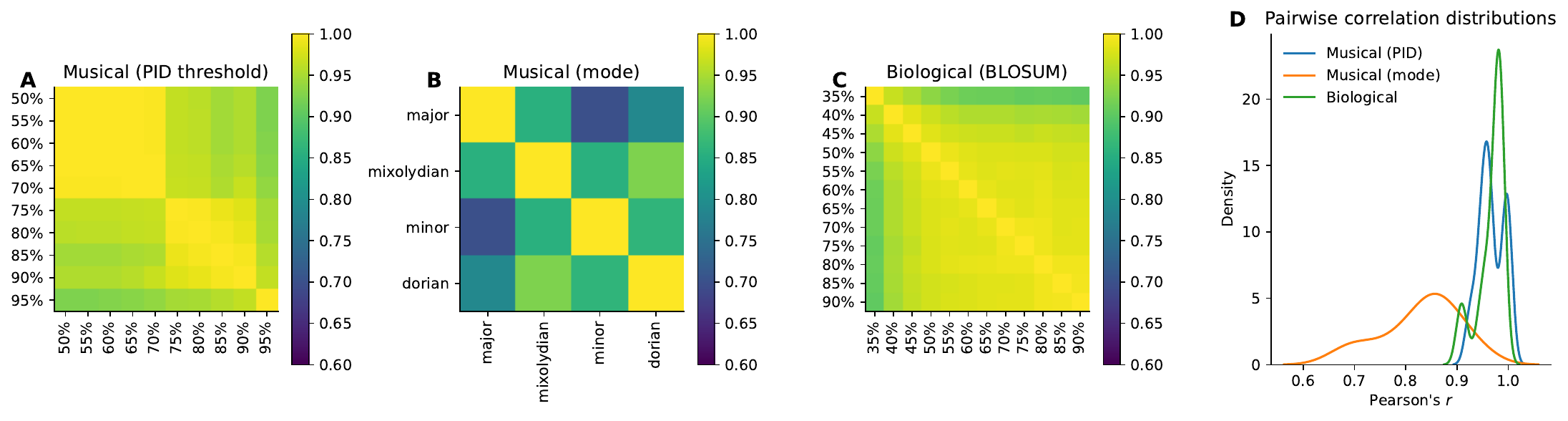}
\caption{\label{fig:4}
  \textbf{Pairwise correlations between substitution matrices.}
  A: Correlations between TheSession log-odds substitution matrices computed
  at different PID thresholds (50\%--95\%).
  B: Correlations between mode-specific substitution matrices (major, minor,
  mixolydian, dorian), computed at $\pid \geq \SI{85}{\%}$.
  C: Correlations between biological substitution matrices (BLOSUM35--BLOSUM90),
  which are constructed at different levels of sequence divergence.
  D: Distribution of off-diagonal pairwise correlations for each group.
  Mode-specific matrices are considerably less similar to each other
  (Pearson's $r \sim 0.6$--$0.8$) than BLOSUM matrices are
  (Pearson's $r > 0.9$), indicating that musical modes differ in their
  substitution patterns more than biological sequences differ across
  divergence levels.
}
\end{figure*}

  A key question is whether mode-specific substitution matrices are
  sufficiently distinct from one another to justify separate treatment,
  and how this inter-matrix variability compares with the variability
  seen among the BLOSUM family of biological substitution matrices.
  BLOSUM matrices are constructed at different levels of sequence
  divergence -- BLOSUM35 is derived from weakly similar sequences and
  BLOSUM90 from closely similar ones -- yet all members of the family
  are highly mutually correlated.
  To make the comparison, we computed pairwise Pearson correlations
  between log-odds matrices in three groups: (i) TheSession matrices
  computed at PID thresholds ranging from \SIrange{50}{95}{\%};
  (ii) the four mode-specific matrices at a fixed threshold of
  $\pid \geq \SI{85}{\%}$; and (iii) BLOSUM35 through BLOSUM90.
  Supplementary \fref{fig:4} shows that the musical matrices are highly self-consistent
  across PID thresholds (panel A), paralleling the behaviour of BLOSUM
  matrices (panel C), whereas the mode-specific matrices are considerably
  less similar to each other (panel B; Pearson's $r \sim 0.6$--$0.9$)
  than BLOSUM matrices are ($r > 0.9$; panel D).
  This confirms that tonality imposes stronger constraints on substitution
  patterns than sequence divergence does in proteins.

  \clearpage

\section{Position-dependent substitution rates for individual meters}
\begin{figure*}[h!]
\centering
\includegraphics[width=0.95\textwidth]{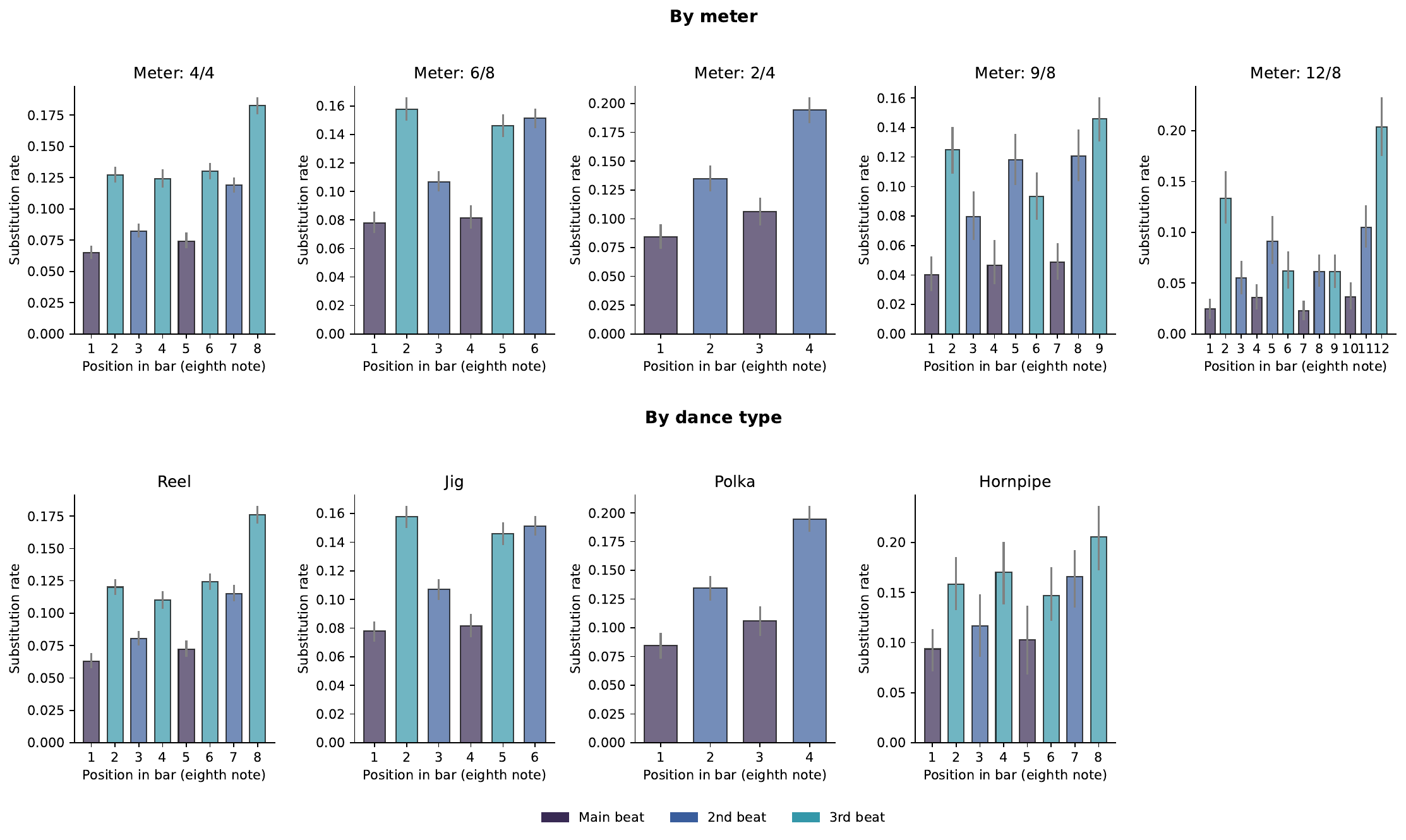}
\caption{\label{fig:5}
  \textbf{Substitution rates vary with metrical position across meters and dance types.}
  Top row: substitution rate vs position within measures for each meter
  (4/4, 6/8, 2/4, 9/8, 12/8).
  Bottom row: substitution rate vs position within measures for each
  dance type (reel, jig, polka, hornpipe).
  Positions are coloured according to metrical hierarchy (\eg, the first
  note in a measure is more important than the second).
  Whiskers show bootstrapped \SI{95}{\%} confidence intervals.
  All rates are calculated for pairs with $\pid \geq \SI{85}{\%}$.
}
\end{figure*}

  In the main text we show that substitution rates vary systematically
  with within-measure position for tunes in 4/4 and 6/8 meter, and that
  this variation closely tracks metrical hierarchy.
  To assess the generality of this finding we repeated the analysis for
  all five meters represented in sufficient numbers in TheSession (4/4,
  6/8, 2/4, 9/8, 12/8) and for each of the four main dance types (reel,
  jig, polka, hornpipe).
  For each group, substitution rates were computed at all within-bar
  positions corresponding to integer multiples of an eighth note,
  following the procedure described in the main text, using all pairs
  with $\pid \geq \SI{85}{\%}$.
  Positions were coloured by their metrical hierarchy level.
  Supplementary \fref{fig:5} shows that the pattern identified in the main text --
  lower substitution rates at metrically strong positions -- is consistent
  across all meters and dance types tested, indicating that the coupling
  between metrical hierarchy and evolutionary constraint is a robust
  feature of this musical tradition rather than an artefact of a
  particular meter or dance class.

  \clearpage

\section{Covariance analysis for individual meters}
\begin{figure*}[h!]
\centering
\includegraphics[width=0.95\textwidth]{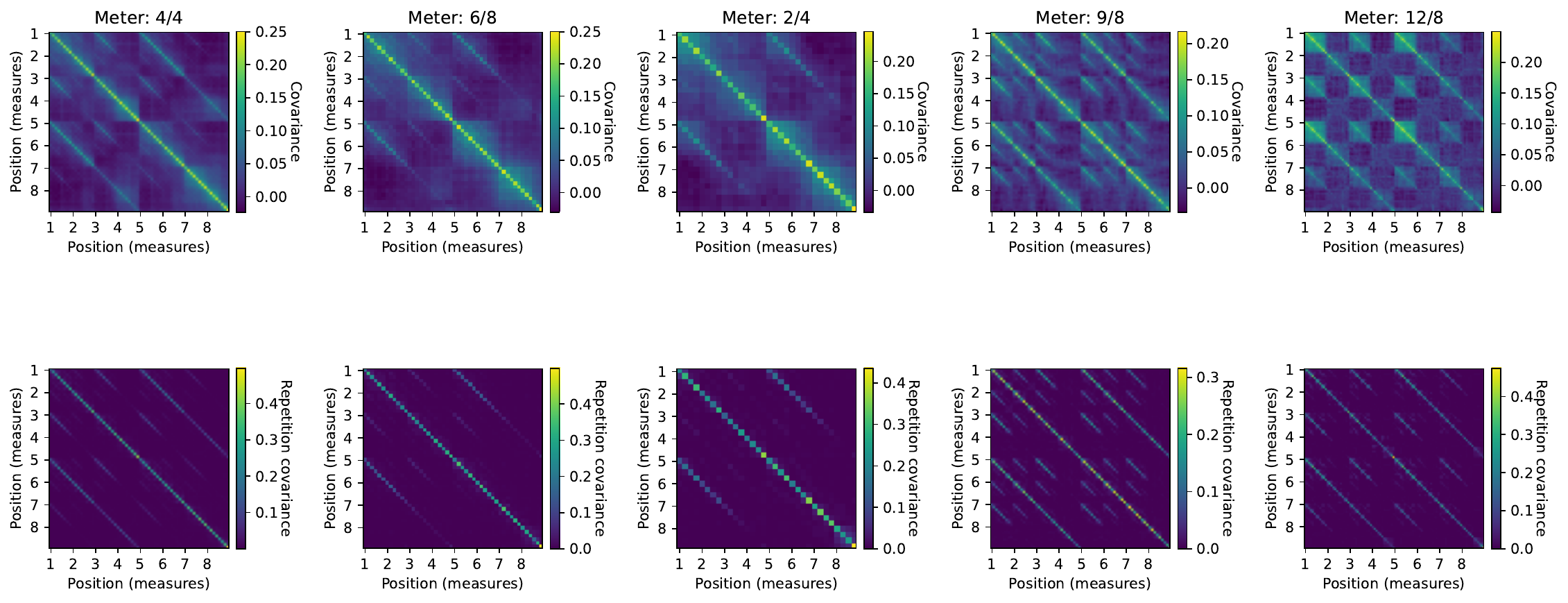}
\caption{\label{fig:6}
  \textbf{Sequence covariance and repetition covariance across all meters.}
  Top row: position-position covariance for tunes grouped by meter
  (4/4, 6/8, 2/4, 9/8, 12/8).
  Bottom row: the component of covariance attributable to repetition
  constraints -- where positions $i$ and $j$ are the same within two
  tunes, but different across them.
  Each cell in the matrix corresponds to a pair of eighth-note positions
  within an 8-bar window.  Axis tick labels indicate measure number.
}
\end{figure*}

  The covariance analysis presented in the main text focuses on 4/4 tunes,
  which constitute the majority of the dataset.
  To determine whether the periodic long-range covariance structure and
  its attribution to repetition constraints are specific to 4/4 meter or
  hold more generally, we repeated both analyses for all five meters in
  the dataset (4/4, 6/8, 2/4, 9/8, 12/8).
  For each meter we computed the position-position covariance matrix over
  the first 8 measures of each part at $\pid \geq \SI{85}{\%}$, as well as
  the repetition-covariance matrix, which isolates pairs of positions that
  carry the same pitch within a tune but differ across variants (see main
  text for methodology).
  Supplementary \fref{fig:6} shows that the periodic covariance structure is present
  in all five meters, and in each case the repetition component reproduces
  the dominant long-range periodicities.
  Note that there are differences, which reflect the known differences in tune
  part structure -- slip jigs (9/8) and slides (12/8), which have 4-part structure,
  repeat at shorter periodicities than the other tunes which have 8-part structure.
  This confirms that the conservation of repeated melodic phrases is a
  general organisational principle of Irish dance tunes, not an artefact
  of the 4/4 meter examined in the main text.

% Bibliography
\bibliography{TheSessionEvo}
\bibliographystyle{unsrtnat}